\documentclass[aip,pof,11pt]{revtex4-2}
\usepackage{subcaption}
\usepackage{graphicx}

\usepackage[english]{babel}
\usepackage{dcolumn}
\usepackage{bm}
\usepackage{xcolor, soul}
\usepackage{amsfonts}
\usepackage{booktabs}
\usepackage{multirow}
\usepackage{siunitx}
\usepackage{ragged2e}
\usepackage[colorlinks,citecolor=red,urlcolor=blue,bookmarks=false,hypertexnames=true]{hyperref} 
\usepackage{float}
\usepackage{xcolor}
\usepackage{array}
\usepackage{setspace}
\usepackage{lipsum}
\usepackage[font=small]{caption}
\usepackage{epstopdf}

\newcommand{\HM}[1]{{\color{black}{#1}}}

\newcolumntype{P}[1]{>{\centering\arraybackslash}p{#1}}
\newcolumntype{M}[1]{>{\centering\arraybackslash}m{#1}}
\usepackage{setspace}
\usepackage{amsmath}
\usepackage{parskip}

\begin{document}
\title{Magnetically Assisted Separation of Weakly Magnetic Metal Ions in Porous Media. Part 2: Numerical Simulations}

\author{Muhammad Garba}
\affiliation{Department of Chemical and Biomedical Engineering, FAMU-FSU College of Engineering, Tallahassee, FL, 32310, USA}
\affiliation{Center for Rare Earths, Critical Minerals, and Industrial Byproducts, National High Magnetic Field Laboratory, Tallahassee, FL 32310, USA}

\author{Alwell Nwachukwu}
\affiliation{Department of Chemical and Biomedical Engineering, FAMU-FSU College of Engineering, Tallahassee, FL, 32310, USA}
\affiliation{Center for Rare Earths, Critical Minerals, and Industrial Byproducts, National High Magnetic Field Laboratory, Tallahassee, FL 32310, USA}
\author{Jamel Ali}
\affiliation{Department of Chemical and Biomedical Engineering, FAMU-FSU College of Engineering, Tallahassee, FL, 32310, USA}
\affiliation{Center for Rare Earths, Critical Minerals, and Industrial Byproducts, National High Magnetic Field Laboratory, Tallahassee, FL 32310, USA}
\author{Theo Siegrist}
\affiliation{Department of Chemical and Biomedical Engineering, FAMU-FSU College of Engineering, Tallahassee, FL, 32310, USA}
\affiliation{Center for Rare Earths, Critical Minerals, and Industrial Byproducts, National High Magnetic Field Laboratory, Tallahassee, FL 32310, USA}
\author{Munir Humayun}
\affiliation{Center for Rare Earths, Critical Minerals, and Industrial Byproducts, National High Magnetic Field Laboratory, Tallahassee, FL 32310, USA}
\affiliation{Department of Earth, Ocean and Atmospheric Science, Florida State University, Tallahassee, FL 32304, USA.}

\author{Hadi Mohammadigoushki}
\thanks{Corresponding author}\email{hadi.moham@eng.famu.fsu.edu}
\affiliation{Department of Chemical and Biomedical Engineering, FAMU-FSU College of Engineering, Tallahassee, FL, 32310, USA}
\affiliation{Center for Rare Earths, Critical Minerals, and Industrial Byproducts, National High Magnetic Field Laboratory, Tallahassee, FL 32310, USA}

\date{\today}

\begin{abstract}
We present a numerical investigation of the magnetophoresis of metal ions in porous media under static, nonuniform magnetic fields. The multiphysics simulations couple momentum transport, mass diffusion, and magnetic field equations, with the porous medium modeled using two distinct approaches: a Stokes-based formulation incorporating effective diffusivity, and a Brinkman-based formulation that explicitly accounts for permeability and medium-induced drag. Comparison with recent experimental data\ [Nwachuwku \textit{et al. Submitted}, 2025] reveals that the Stokes model partially fails to capture key trends, while the Brinkman model, with permeability accurately reproduces observed transport behavior on various porous media. Our simulations predict that both paramagnetic (MnCl$_2$) and diamagnetic (ZnCl$_2$) ions may form field-induced clusters under magnetic gradients over a range of concentrations of 1mM-100mM and magnetic field gradients of up to 100 T$^2$/m. The dominant driving force is found to be the magnetic gradient (Kelvin) force, while the paramagnetic force from concentration gradients contributes minimally. In binary mixtures, hydrodynamic interactions between paramagnetic and diamagnetic clusters significantly alter transport dynamics. Specifically, paramagnetic clusters can pull diamagnetic clusters along the magnetic field gradient, enhancing diamagnetic migration and suppressing the motion of paramagnetic species. These findings highlight the importance of porous media modeling and interspecies interactions in predicting magnetophoretic transport of ionic mixtures.

\end{abstract}
\maketitle
\section{Introduction}
Separation processes are a foundational aspect of chemical engineering~\cite{seader2011separation,constable2019sustainable}, accounting for up to 15\% of global energy consumption~\cite{humphrey1997separation,MaterialsSeparationTechnologies2005,sholl2016seven}. These processes enable the isolation, purification, and recovery of target components from complex mixtures, and are integral to host of sectors such as chemical manufacturing, environmental remediation, water treatment, and biotechnology~\cite{osti_131708}. Conventional separation techniques such as distillation, crystallization, and/or liquid–liquid extraction are technologically mature but often associated with high energy demands, environmental impacts, and operational constraints~\cite{seader2011separation,seider2016product}. Their reliance on elevated temperatures, pressures, or chemical additives frequently results in the generation of secondary waste streams and limits their compatibility with emerging sustainability-oriented processes~\cite{constable2019sustainable}. Consequently, there is growing interest in alternative separation technologies that offer improved energy efficiency with reduced environmental impact~\cite{constable2019sustainable}. A broad range of advanced particle manipulation techniques has emerged in recent years, including electrophoresis~\cite{rana2023electrophoresis}, dielectrophoresis~\cite{luna2023enhancement,kim2018dielectrophoresis}, thermophoresis~\cite{seidel2013microscale}, acoustophoresis~\cite{petersson2007free}, optical trapping~\cite{neuman2004optical}, and magnetophoresis~\cite{zhu2010particle,munaz2018recent,khan2025magnetophoresis,rassolov2025magnetophoresis}. Among these methods, magnetophoresis (or transport of magnetic particles under non-uniform magnetic fields) requires a low energy demand, operational simplicity, and system scalability making it a promising candidate for applications in renewable energy-driven separation systems and emerging green technologies~\cite{zhu2010particle, pulko2014magnetic, zborowski1999magnetophoresis}. Magnetophoresis has already found application in drug delivery~\cite{furlani2010magnetic}, bioseparation and medical imaging~\cite{ayansiji2020constitutive}, and the removal of toxic heavy metals from wastewater streams~\cite{leong2020unified}.\par 

Magnetophoresis relies on the differential response of materials to an applied inhomogeneous magnetic field~\cite{fujiwara2001separation, pulko2014magnetic, franczak2016magnetomigration, shi2016simulation}. Magnetic materials may be classified as paramagnetic or diamagnetic, depending on their intrinsic magnetic susceptibility. Paramagnetic species are attracted toward regions of higher magnetic field intensity and are characterized by positive magnetic susceptibility, whereas diamagnetic particles exhibit negative magnetic susceptibility and migrate toward lower field regions~\cite{munaz2018recent,iacovacci2016magnetic}. Most existing studies on magnetophoresis focus on nanoparticles, colloids, or ferrofluids~\cite{Leong2020Langmuir,Leong2015SoftMatter,yavuz2006low,lim2014challenges,schaller2008motion,de2008low,ge2017magnetic,watson1975theory}, while comparatively fewer investigations have examined the magnetophoresis of dissolved metal ions. Metal ions, being significantly smaller, may exhibit distinct transport behaviors influenced by solvation effects, diffusion, and interactions with the surrounding matrix. Metal ions commonly exist as mixtures in many industrial and environmental contexts, necessitating their selective separation and purification. Examples include wastewater treatment, critical metal for resource recovery~\cite{wangselective,benhal2025dynamics}, and analytical chemistry~\cite{iranmanesh2017magnetic}. These needs have driven growing interest in magnetophoresis as a viable technique for metal ion separation. To date, the magnetophoresis of paramagnetic and diamagnetic metal ions has been studied in both fluidic systems~\cite{benhal2025dynamics,pulko2014magnetic, yang2012enrichment, yang2014magnetic, rodrigues2017magnetomigration, rodrigues2019effect, kolczyk2016holmium, kolczyk2016rareearth, lei2017evaporation} and porous media~\cite{Nwachuwku2025,fujiwara2001separation, fujiwara2004movement,franczak2016magnetomigration,fujiwara2006movement}. \par 

\citeauthor{yang2012enrichment} investigated the enrichment of paramagnetic MnSO$_4$ in solution under a non-uniform magnetic field generated by a cylindrical permanent magnet placed at the top of the container for concentrations ranging from 0.1 to 1.0 M~\cite{yang2012enrichment}. A maximum enrichment of about 2\% was observed in regions where the magnetic field gradient was high. The degree of enrichment was found to increase with both initial concentration and time~\cite{yang2014magnetic}. Similar behavior was reported for GdCl$_3$, but not for CuSO$_4$, due to its significantly lower magnetic susceptibility~\cite{yang2012enrichment}. In a follow-up study, \citeauthor{yang2014magnetic} concluded that the configuration of the magnetic field also plays a critical role in the enrichment process~\cite{yang2012enrichment}. \citeauthor{pulko2014magnetic} extended this work to lanthanide salts such as DyCl$_3$, and interestingly found that, despite DyCl$_3$ having a magnetic susceptibility three times greater than MnSO$_4$, its enrichment was lower~\cite{pulko2014magnetic}. While magnetic field gradients are conventionally considered essential for magnetophoresis~\cite{yang2012enrichment, yang2014magnetic}, \citeauthor{rodrigues2019effect} showed that even in initially uniform magnetic field, concentration gradients, originating from evaporation or temperature variations, can initiate magnetically driven transport~\cite{rodrigues2019effect, rodrigues2017magnetomigration}. Notably, diamagnetic ions have not demonstrated significant enrichment or depletion under similar conditions~\cite{rodrigues2017magnetomigration, rodrigues2019effect}. More recently, Benhal et al.~~\cite{benhal2025dynamics} investigated the separation of transition metal ions using a high gradient magnetic separator. Their results showed that paramagnetic ions were effectively captured by magnetic field gradients near magnetized wires, whereas diamagnetic ions exhibited negligible interaction with the magnetic mesh~\cite{benhal2025dynamics}. In fluidic systems, magnetic species are simultaneously influenced by magnetic, diffusive, viscous, and/or gravitational forces. This complexity presents challenges in quantitatively characterizing magnetophoretic transport. To address this, porous media have been widely used as a model system to minimize convective effects and enable controlled investigation of the interplay between magnetic forces and diffusion~\cite{Nwachuwku2025,fujiwara2001separation, fujiwara2004movement,franczak2016magnetomigration,fujiwara2006movement,Rassolov2024magnetophoresis}.\par 

In porous media, \citeauthor{fujiwara2004movement} investigated the migration of transition metal ions spotted onto silica gel under a magnetic field gradient generated by a superconducting magnet~\cite{fujiwara2004movement}. Paramagnetic ions migrated toward regions of higher field strength, at rates much higher than expected dependent on their susceptibility and concentration, while diamagnetic ions remained stationary~\cite{fujiwara2001separation, fujiwara2004movement}. \citeauthor{franczak2016magnetomigration} demonstrated that paramagnetic ions accumulate near regions of high magnetic field gradients, whereas diamagnetic ions are depleted from these regions~\cite{franczak2016magnetomigration}. Notably, their findings indicated that magnetic susceptibility had limited influence on the extent of magnetophoresis. However, interpretation was complicated by metal ion adsorption to their porous media~\cite{fujiwara2004movement,franczak2016magnetomigration}. To address this, our group recently conducted a systematic investigation of the magnetophoresis of paramagnetic MnCl\textsubscript{2} and diamagnetic ZnCl\textsubscript{2} in silica gel under a non-uniform magnetic field generated by a permanent magnet~\cite{Nwachuwku2025}. We observed enrichment of MnCl$_{2}$ near the magnet surface and depletion of ZnCl$_2$ from this region. In binary mixtures, both ions migrated toward areas of high magnetic field gradients, and \HM{ the results imply potential formation of magnetic field-induced ion clusters~\cite{Nwachuwku2025}. It should be noted that to date, there is no direct experimental observation of magnetically induced ion clusters. Nevertheless, previous studies such as those by \citeauthor{georgalis2000cluster}~\cite{georgalis2000cluster} have reported the formation of solvated ion and molecular clusters in concentrated electrolyte solutions using dynamic light scattering. Similarly, \citeauthor{komori2023cluster}~\cite{komori2023cluster}, \citeauthor{safran2023scaling}~\cite{safran2023scaling} and \citeauthor{bian2011ion} provide further evidence of nanoscale ionic aggregation in the absence of external magnetic fields. More recently, \citeauthor{dinpajooh2025magnetic}~\cite{dinpajooh2025magnetic}introduced the term nanoscale domains to describe correlated paramagnetic ion assemblies whose collective magnetic dipole interactions can influence aggregation at the nanometer scale under the influence of an external magnetic field.}\par 

Although several experimental studies have investigated the magnetophoresis of metal ions in porous media, a comprehensive theoretical framework that captures the underlying transport mechanisms remains underdeveloped. In particular, while experimental evidence points to the formation of ion clusters under magnetic fields, the dynamics governing their formation and evolution over time are not well understood. The magnetic potential energy $E_m$ acting on a magnetized species in an external magnetic field is given by~\cite{zborowski1999magnetophoresis, ayansiji2020constitutive, svoboda2004magnetic, coey2007magnetic}:
$E_m = -\frac{V}{2\mu_0} \Delta\chi_m c\mathbf{B}^2,$ where $\Delta \chi_m$ is the molar magnetic susceptibility difference between the species and the surrounding medium, $\mathbf{B}$ is the magnetic flux density, $V$is the particle volume, $c$ is the concentration, $\mu_0 = 4\pi \times 10^{-7}$is the vacuum permeability, and $R$ is the effective particle radius. Under isothermal conditions, the resulting magnetic force is given by: 
\begin{equation}\label{magneticforce}
\mathbf{F}_m =  -\nabla E_m = \frac{V \Delta\chi_m}{2\mu_0} \nabla(c \mathbf{B}^2).
\end{equation}
Therefore, magnetic force has two distinct contributions: one arising from gradients in the magnetic field (commonly referred to as the Kelvin force) and another from spatial gradients in concentration, referred to here as the paramagnetic force. \HM{The paramagnetic force is also known as Korteweg-Helmholtz force elsewhere~\cite{Butcher2023JPhysChemB}.} The relative importance of these two components remains a subject of debate~\cite{coey2007magnetic, yang2012enrichment, leventis2005demonstration, hinds2001influence, rodrigues2017magnetomigration, rodrigues2019effect}. Some studies argue that the paramagnetic force is negligible in dilute systems or under uniform magnetic fields~\cite{coey2007magnetic, yang2012enrichment}, while others have demonstrated its influence in initiating magnetophoretic transport even in initially homogeneous systems~\cite{leventis2005demonstration, hinds2001influence, rodrigues2017magnetomigration, rodrigues2019effect}. For example, \citeauthor{rodrigues2017magnetomigration} proposed that magnetophoretic motion in their porous gel system was driven in part by concentration gradients induced during the gelation process, suggesting that mechanisms beyond the Kelvin force are relevant~\cite{rodrigues2017magnetomigration, franczak2016magnetomigration}. This discrepancy highlights a fundamental knowledge gap, particularly in porous media, where magnetic field and concentration gradients are inherently coupled and experimentally difficult to decouple. Furthermore, the influence of hydrodynamic drag on magnetophoretic transport in porous systems has not been systematically examined. Drag forces can become significant in confined porous geometries, where increased surface interactions and restricted flow pathways may alter transport behavior and influence magnetophoretic dynamics.\par 

This study presents multiphysics simulations of paramagnetic and diamagnetic ion magnetophoresis in porous media under non-uniform magnetic fields. The model builds upon and extends the framework developed by Rassolov et al.~\cite{Rassolov2024magnetophoresis}. \HM{We should note that in our previous model~\cite{Rassolov2024magnetophoresis}, we did not account for the influence of the paramagnetic force on the magnetophoresis of metal ions in porous media. In this study, we investigate how incorporating this contribution affects the overall magnetophoretic motion within porous structures.} The simulations are systematically compared with experimental results reported in Part (I) of this series~\cite{Nwachuwku2025}, which characterized the magnetophoresis of MnCl$_2$ and ZnCl$_2$ in silica gel. Our simulations aim to address several open questions that remain unresolved by prior experiments alone. Specifically, we investigate the formation and evolution of magnetic field-induced ion clusters, probing their dynamics over time. We further quantify the relative contributions of the Kelvin force and the paramagnetic concentration-driven force to overall ion transport. Additionally, we examine the influence of hydrodynamic drag in porous media by incorporating both Stokes and Brinkman formulations, providing insight into how confinement and pore-scale resistance impact magnetophoretic motion. Finally, the model is applied to binary mixtures of paramagnetic and diamagnetic ions to investigate their transport behavior and to elucidate the underlying mechanisms governing competitive and cooperative ion-ion interactions during magnetically-induced separation.\par 
\section{Model and Governing equation}
\begin{figure}[hthp]
    \centering
    \subfloat{{\includegraphics[width=\linewidth]{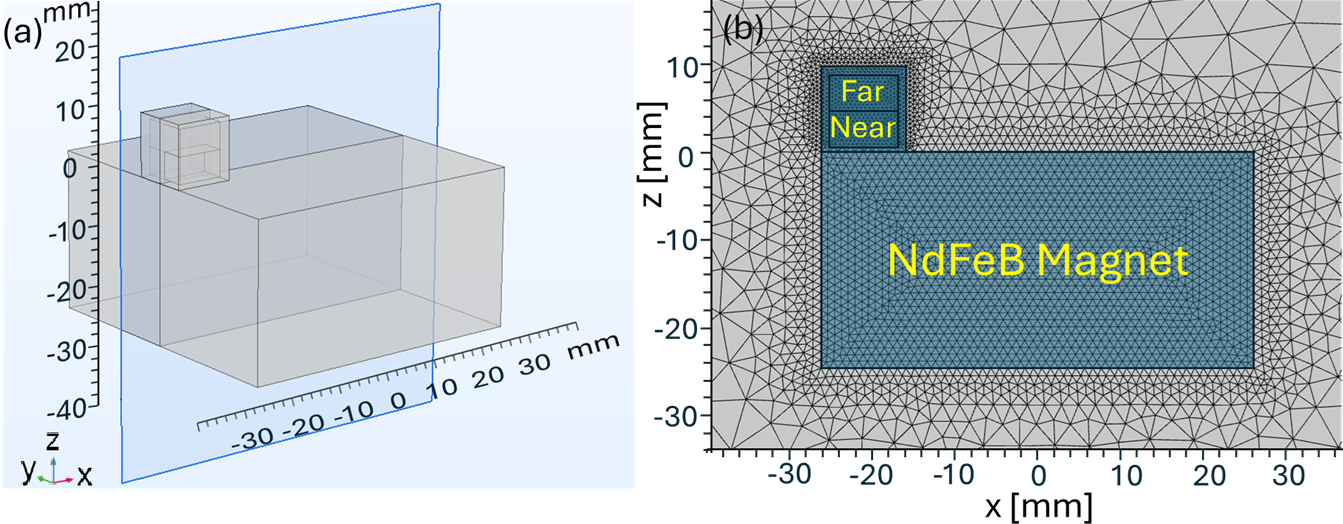}}}
   
        \caption{Schematic representation of the experimental set-up used in part (I) of this series (a) and a cross-section of the domain showing the NdFeB permanent magnet and the cell containing porous media along with the representative mesh used in 2D simulations (b). }
    \label{Model figure}
\end{figure}

To simulate the magnetophoresis experiments described in Part (I) of this series, a closed computational domain is constructed, representing a cell positioned above a Neodymium (NdFeB) permanent magnet, as illustrated in Fig.~(\ref{Model figure}). The large solid block denotes a 50.8$\times$50.8$\times$25.4 mm$^3$ permanent magnet, while the smaller block atop the magnet (10$\times$10$\times$10 mm$^3$) represents the cell containing a porous medium saturated with an initially homogeneous solution of metal ions consistent with experimental parameters. To investigate the spatial enrichment and depletion of metal ions within the domain, the cell is further subdivided into two equal regions designated as the near and far regions relative to the magnet surface, and shown in Fig.~\ref{Model figure}(b). \par 
The high computational cost associated with full 3D simulations renders a comprehensive parametric study impractical within a reasonable timeframe. Therefore, a 2D computational domain, based on a representative cross-sectional slice of the 3D geometry, is primarily employed to simulate the magnetophoresis process. The model equations presented below were numerically solved using the finite element method in COMSOL Multiphysics 6.1. As shown in Fig.~\ref{Model figure}(b), the cell was discretized with 4,042 triangular mesh elements, with a maximum and minimum element size of 0.25 mm and 0.01 mm respectively. The permanent magnet was meshed using 3,306 triangular mesh elements with a maximum element size of 1 mm. The external region encompassing both the cell and the magnet was discretized with 3,996 triangular elements with a maximum and minimum element size of 20.4 mm and 0.0914 mm respectively. The entire magnetic field domain spans a 12$\times$7 in$^2$ rectangular region, comprising a total of 11,344 triangular mesh elements. Simulating the magnetophoresis of metal ions within the cell requires solving both the magnetic field, momentum and mass transport equations, which will be discussed in the following sections.\par 

\subsection{Static Magnetic Field } \label{MagField}
The magnetic field around the permanent magnet was first modeled to obtain the magnetic flux density \textbf{B} [T] and then further processed to obtain the inner product of magnetic flux density and its gradient ($\mathbf{B}\cdot\nabla \mathbf{B}$) [T$^2$/m].  The relationship between magnetic field, \textbf{H} and the scalar potential, V$_m$ for the entire domain is given by:
\begin{equation}\label{mf1}
    \mathbf{H}=-\nabla V_m
\end{equation}
Also, the relationship between the magnetic field strength, \textbf{H} and the magnetic flux density, \textbf{B} outside the magnet is expressed using Gauss's law as:
\begin{equation}\label{mf2}
\begin{split}
    & \nabla\cdot \mathbf{B}=0; \\ 
    &\mathbf{B}=\mu_0\mu_r \mathbf{H}
\end{split}
\end{equation}

where $\mu_r$ is the relative permeability, which is assumed to be approximately 1 within the external magnetic field. However, inside the permanent magnet, $\mathbf{B} = \mu_0\mu_{rec} \mathbf{H} + \mathbf{B}_r$, where $\mathbf{B}_r$ is the remanence flux density, and $\mu_{rec}$ is the recoil permeability. In these simulations, the magnet is modeled using the N52 (sintered NdFeB) material in COMSOL; therefore, $\mathbf{B}_r = 1.44~\si{T}$ and $\mu_{rec} = 1.05$. Finally, a zero magnetic flux condition ($\mathbf{n} \cdot \mathbf{B} = 0$) was imposed far away from the magnet.\par

\subsection{Transport of Metal Ions in Porous Media}
Metal ion mass transport in the computational domain could be quantified using the drift-diffusion equation for porous media as~\cite{Rassolov2024magnetophoresis}: 
\begin{equation}\label{drift-diffusion}
\begin{split}
    & \varepsilon\frac{\partial c}{\partial t}+\nabla \cdot \mathbf{N}=0 \\
    & \mathbf{N}=-D_{e}\nabla c+c\mathbf{v_{mig}}
\end{split}
\end{equation}
where $c$ is the concentration of metal ion, $t$ is time, $D_{e}$ is the effective diffusion coefficient of metal ions in the porous medium, $\mathbf{v_{mig}}$ is the drift velocity, and $\varepsilon$ is the porosity of the media. In addition, $\mathbf{N}$ is the total molar flux of species in the domain that includes both diffusion and convective fluxes. The drift velocity $\mathbf{v_{mig}}$ is determined by balancing the net magnetic forces acting on the metal ion with the drag force exerted by the surrounding medium, while neglecting inertial, gravitational, and buoyancy forces~\cite{Rassolov2024magnetophoresis}. \HM{In addition, hydrodynamic fluid convection can be neglected because metal ions are transported through the porous matrix. Furthermore, in this model, the solution of metal ion is considered to be ideal and therefore, diffusion is Fickian.} \par 
The porous medium used in the experimental section of this study consists of a packed bed of solid silica particles, forming an interconnected network of voids through which fluid and suspended metal ions can migrate. The transport of metal ions through such a medium is governed by complex interactions among the fluid phase, the solid matrix, and the externally applied magnetic field. In this paper, we investigate the influence of the porous medium on ion transport through two complementary modeling approaches. In the first, simplified approach, the porous medium is treated as a homogeneous continuum medium, where self-diffusion of metal ions is characterized by an effective diffusion coefficient, 
$D_e$, which accounts for the geometric constraints imposed by the porous structure.\HM{ Specifically, 
$D_e$ is estimated using the well-known Millington and Quirk empirical model $D_e={\varepsilon^{4/3}}D_f$, a widely used empirical relation that accurately estimates diffusivity in porous media.}~\cite{millington1961permeability}. In here $D_f$ denotes the free self-diffusion coefficient of metal ions in solutions. In this framework, the magnetophoretic motion of ions is governed by a steady-state force balance between the magnetic force and viscous drag, expressed via Stokes' law ($\mathbf{F_{ds}} = 6\pi\eta R_{s}\mathbf{v_{mig}}$ \cite{zborowski1999magnetophoresis,Rassolov2024magnetophoresis}) and is expressed as:  
\begin{equation}\label{stoke's eqn}
   \mathbf{F_m}-\mathbf{F_{ds}}=0;\qquad \mathbf{v_{mig}}=\frac{R^3 \Delta \chi_m}{9\mu_0\eta R_{s}} \nabla {(c\mathbf{B}^2)},
\end{equation}
where $\eta$ is the solution viscosity and $R_s$ is the hydrodynamic radius of the metal ion and is assumed to be equal to the hydration radius of the metal ions $ R = R_{s}$. \HM{In addition, $\Delta \chi_m$ is the difference between magnetic susceptibility of the metal ion and the solvent (water in this case). The molar magnetic susceptibility $\chi_m$ of MnCl$_2$ and ZnCl$_2$ considered in our model are extracted from CRC handbook and shown in Table~\ref{Mag_suscep} in cgs unit~\cite{haynes2016crc}. The molar magnetic susceptibility of a solute is an intrinsic property determined by the electronic configuration of the metal ion, independent of whether the solvent is in bulk or confined within pores. In a saturated porous medium, the local chemical environment (e.g., hydration and coordination) remains essentially the same unless strong adsorption occurs. Since metal ion adsorption to porous media was minimized by lowering the pH, as confirmed by control experiments showing no change in ion concentration without a magnetic field\cite{Nwachuwku2025}, the intrinsic molar magnetic susceptibility of the metal ions remains unaffected by the porous medium. \par}
\begin{table}[hthp]
    \centering
    \caption{Molar Magnetic Suceptibility of metal ions used in this study.}
     \label{Mag_suscep}
        \begin{tabular}{|c|c|}
    \hline
        \textbf{Solute} & \textbf{$\chi_m$/10$^{-6}$ [cm$^3$ mol$^{-1}$}]\cite{haynes2016crc} \\
        \hline
       MnCl$_2$  &  14350 \\
       ZnCl$_2$  & -55.33  \\
       \hline
    \end{tabular}

\end{table} 

While the first approach offers a simplified yet effective framework for capturing magnetophoresis of metal ions in porous media~\cite{Rassolov2024magnetophoresis}, it is well established that the drag experienced by species in a porous network can significantly deviate from classical Stokes' drag~\cite{deo2010drag, ramalakshmi2019drag}. This deviation arises from the confined and spatially heterogeneous structure of porous media, which imposes additional hydrodynamic resistance on migrating species. To account for these complexities, a second, more refined modeling approach employs the Brinkman equation. This equation extends Darcy’s law by incorporating viscous shear effects, effectively bridging the gap between Darcy flow, appropriate for highly permeable media, and Stokes flow in unbounded fluids. In a seminal work, Brinkman~\cite{brinkman1949calculation} developed a modified Stokes' drag in a porous medium made of spherical particles as: $\mathbf{F_{db}}=6m\pi\eta R\mathbf{v_{mig}}$, where $m$ denotes a correction factor to Stokes' drag, and is given by: 

\begin{equation}\label{brinkman_equation2}
  m = 1 +\lambda R+\frac{\lambda^2R^2}{3};\qquad \lambda=\left(\frac{\eta}{\kappa\eta_e}\right)^\frac{1}{2}; \qquad \eta\approx\eta_e.
\end{equation}

Here $\eta_e$ is the effective viscosity of the fluid, and $\kappa$ is the permeability of the porous media that can be assessed using Carman-Kozeny model \cite{rehman2024comprehensive} as: \begin{equation}\label{Carman-Kozeny}
    \kappa=\Phi^2\frac{d_p^2\varepsilon^3}{180(1-\varepsilon)^2}.
\end{equation}
Here $d_{p}$ is the porous media particle size (diameter) and $\Phi$  is the sphericity factor characterizing the shape of particles that make the porous media ($\Phi = 1$ for a spherical particle and $\Phi < 1$ for non-spherical particles). Note that Brinkman drag reduces to Stokes' drag in the limit of $\kappa \rightarrow \infty$. Therefore, in the second approach, the magnetophoretic velocity of metal ions is given by: \begin{equation}\label{Brinkman's eqn}
   \mathbf{F_m}-\mathbf{F_{db}}=0;\qquad \mathbf{v_{mig}}=\frac{R^3 \Delta \chi_m}{9m\mu_0\eta R_{s}} \nabla {(c\mathbf{B}^2)}.
\end{equation} 

The multiphysics numerical simulations were carried out in a sequential manner. First, the static magnetic field distribution was computed by solving Eqs.~\eqref{mf1} and~\eqref{mf2}. The resulting steady-state magnetic field solution was applied to the drift-diffusion model described by Eq.~\eqref{drift-diffusion} to simulate the migration of metal ions within the porous media.

\section{Results and Discussion}
\subsection{Static Magnetic Field}

Before performing magnetophoresis simulations, it is essential to validate the accuracy of the computed static magnetic field by confirming that it captures the correct spatial distribution and quantitatively agrees with experimental measurements. To this end, we present the spatial profiles of the magnetic flux density, $\mathbf{B}$, and its field gradient, $(\mathbf{B} \cdot \nabla)\mathbf{B}$, within the computational domain. These quantities are critical for accurately resolving the magnetic force experienced by metal ions and for ensuring the fidelity of subsequent transport simulations.\par

\begin{figure}[hthp!]
 \centering

\includegraphics[width=\linewidth]{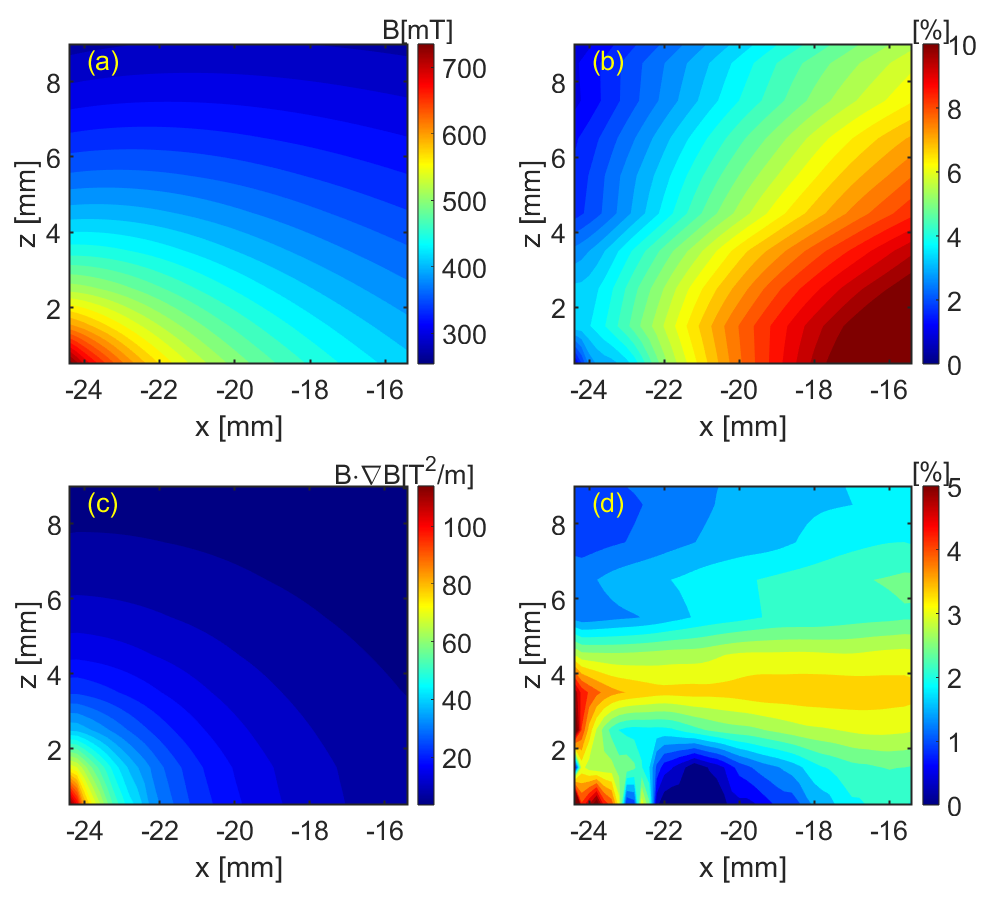}

\caption{(a) Magnetic flux density ($\mathbf{B}$) map calculated by 2D-Simulations. (b) Normalized percentage difference between computed magnetic flux density and the experiments. (c) Calculated magnetic flux density gradient $\mathbf{B}\cdot\nabla \mathbf{B}$ in 2D-simulations. (d) Normalized percentage difference in magnetic field gradients between experiments reported in part (I) and calculations of this study. }
    \label{BBgradB fields}
\end{figure}

Fig.~\ref{BBgradB fields}(a) illustrates the magnetic flux density $\mathbf{B}$ within the simulation domain of the cell computed using equations described in Section~(\ref{MagField}) above. As expected, the field intensity increases near the magnet surface and reaches a maximum value at the magnet's edge. Fig.~\ref{BBgradB fields}(b) shows the relative percentage error between the simulated magnetic flux density and the measured experimental values reported by Nwachuwku et al.~\cite{Nwachuwku2025}. The difference is calculated as: ${| \mathbf{B}_{\text{exp}} - \mathbf{B}_{\text{sim}}|}/{\mathbf{B}_{\text{exp}}} \times 100 [\%]$. Fig.~\ref{BBgradB fields}(b) shows that the percentage deviation is highest (approximately 10\%) near the magnet surface but away from its edge, where magnetic flux density is at its minimum. This discrepancy is mainly attributed to the inherent limitations of the two-dimensional simulation, which does not account for variations in the out-of-plane direction due to boundary effect. Nevertheless, the deviation remains minimal in regions where the magnetic field is strongest, indicating that the simulation adequately captures the key features of the experimental data.\par

More importantly, Fig.~\ref{BBgradB fields}(c) presents the calculated magnetic field gradients, while Fig.~\ref{BBgradB fields}(d) shows the differences between field gradients of simulations and experimental measurements of Nwachuwku et al.~\cite{Nwachuwku2025}. The discrepancy between the simulated and experimentally measured field gradients is found to be less than 5\% across the computational domain. It is worth noting that a portion of this difference arises from the numerical differentiation of experimental data, which tends to amplify measurement noise. Despite this, the close agreement between simulation and experiment confirms the validity of the computed magnetic field. Therefore, the simulated field can be confidently used as input for the subsequent magnetophoresis simulations.\par

\subsection{Magnetophoresis of Single Metal Ions}
\subsubsection{Fixed Size Metal Ion Clusters}

\begin{figure*}[ht]
\centering
{\includegraphics[width=\linewidth]{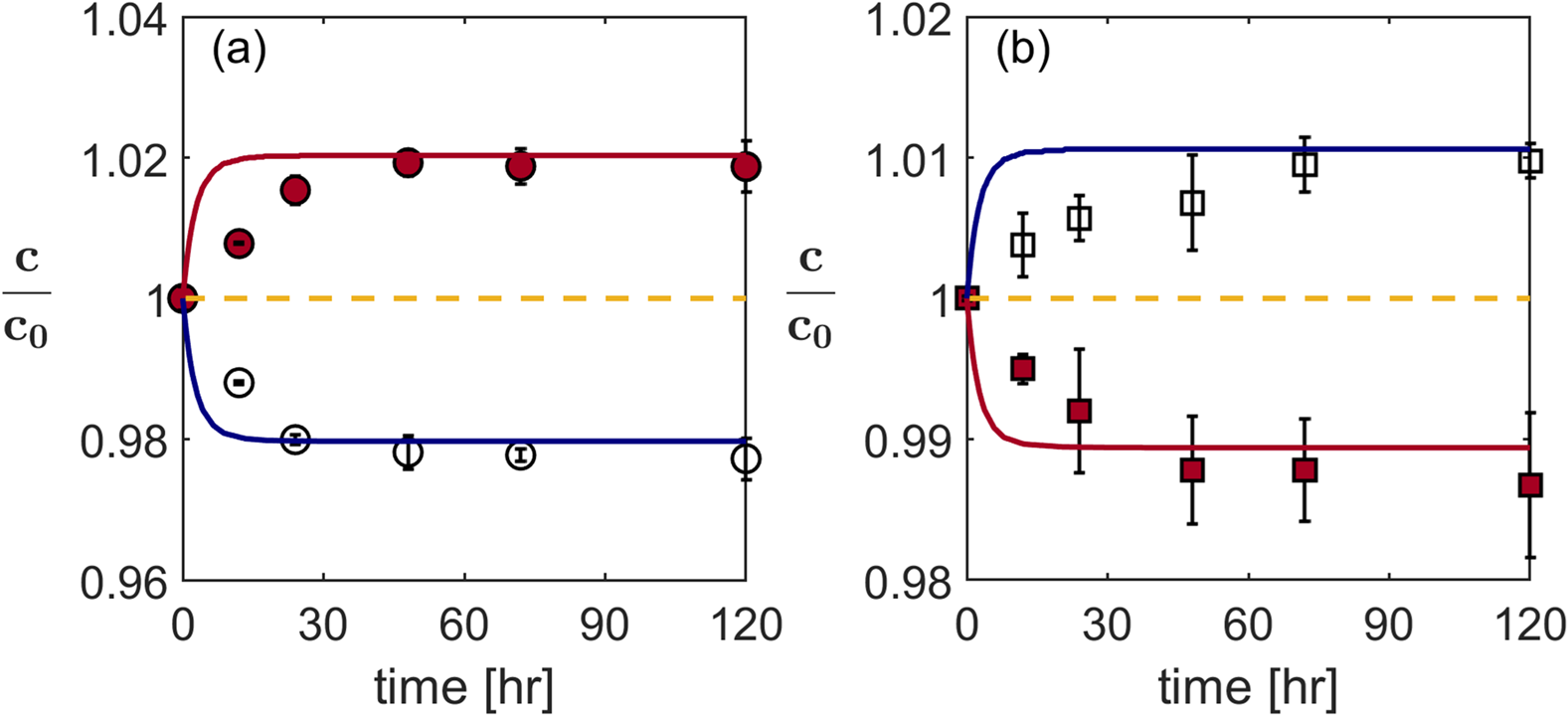}}\\  
{\includegraphics[width=0.8\linewidth]{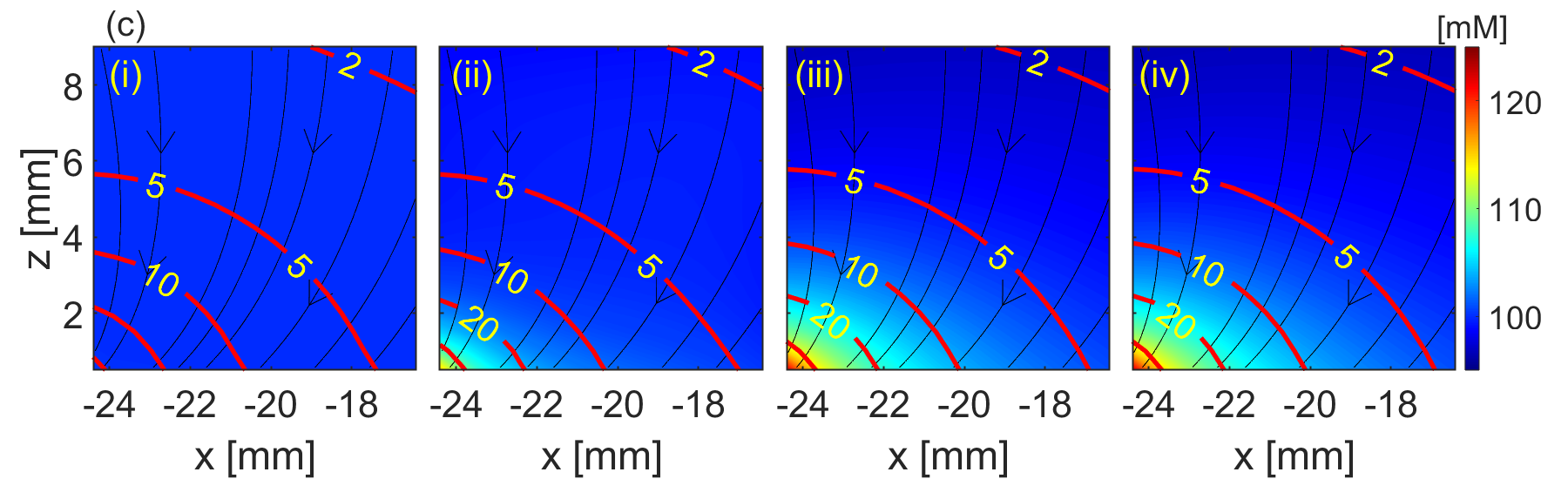}}\\
{\includegraphics[width=0.8\linewidth]{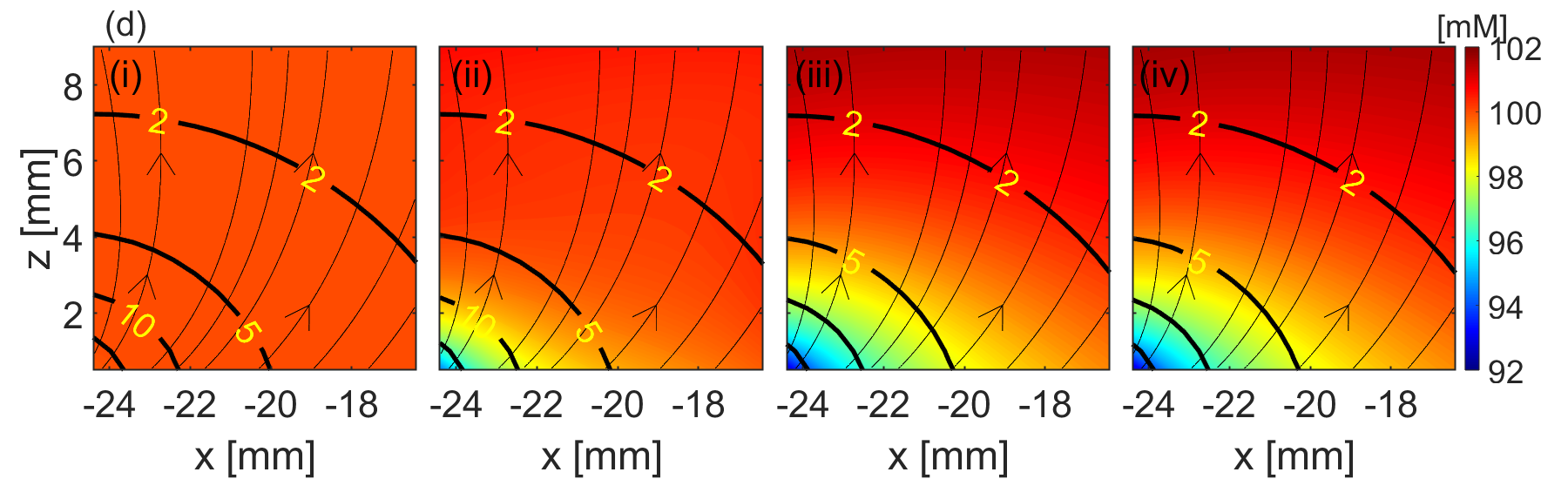}}
 \caption{Temporal evolution of metal ion concentration along with the predictions of the model (solid and dashed curves) for MnCl$_2$ (a) and ZnCl$2$ (b) at an initial concentration of c$_\text{0}$ = 100mM. In part (a) and (b), the filled and open symbols denote the concentrations in the near and far sections, respectively. The dashed line shows the model predictions for metal ion hydration size, and the continuous curves denote the model predictions for a constant cluster size. The spatio-temporal evolution of metal ion concentration along with velocity vectors and velocity contours for MnCl$_2$(c) and ZnCl$_2$ (d) for an initial concentration of 100mM after t = 0 (i), t = 1 (ii), t = 12 (iii), and t = 72 (iv) hours. Velocity labels in (c,d) have a unit of nm/s.}
    \label{constantcluster}
\end{figure*}

Fig.~(\ref{constantcluster}) shows the temporal evolution of paramagnetic and diamagnetic metal ion concentration for an initial concentration of 100mM. As discussed in Part (I) of this study, the paramagnetic MnCl$_2$ exhibits enrichment near the magnet, with an increase of approximately 2–3\%, whereas the diamagnetic ZnCl$_2$ consistently shows depletion in the near region across all tested concentrations. At low electrolyte concentrations (typically $<$100 mM), metal ions in aqueous solutions are generally expected to form ion pairs or small clusters through interactions with counterions. In particular, oppositely charged ions are known to associate into solvent-separated or contact ion pairs, resulting in the formation of hydrated complexes~\cite{conway1981ionic}. \HM{Experimental evidence by \citeauthor{georgalis2000cluster}~\cite{georgalis2000cluster} supports this notion, showing the formation of solvated ions and molecular clusters in concentrated electrolyte solutions using dynamic light scattering under nonmagnetic conditions. These observations highlight that ion aggregation is not limited to dilute regimes but extends into more concentrated systems, where collective solvation effects become significant~\cite{angarita2022ion, jiang2025investigation}.
} Therefore, as an initial step in modeling magnetophoresis under these conditions, metal ions were assumed to form hydrated complexes in aqueous solution, with an effective hydrodynamic radius of approximately $R_s \approx 6 \textup{\AA}$ \cite{persson2024structure}. This value reflects both the ionic core and the primary hydration shell and is consistent with literature estimates for divalent metal cations in aqueous environments. Accordingly, it was adopted as the representative particle size in our simulations to model ion transport. The corresponding model predictions are shown in Fig.~\ref{constantcluster}(a,b) as dashed lines. Under this assumption, the simulations predict negligible magnetophoretic transport, and thus no effective separation. Similar results have been obtained at various initial concentrations (see Fig.~S1, and Fig.~S2 in the supplementary materials). The latter result is in contrast to the experimental observations that show magnetophoretic motion of metal ions.\par 

In fact, these findings are consistent with our recent modeling efforts as well as prior experimental studies reported in the literature~\cite{fujiwara2004movement, chie2003magnetic, franczak2016magnetomigration,Rassolov2024magnetophoresis}. In prior studies, researchers observed significant magnetophoretic transport of a range of metal ions in porous media, and postulated that the application of a magnetic field may induce formation of larger ionic clusters~\cite{fujiwara2006movement, franczak2016magnetomigration}. Such magnetic field-induced clustering could increase the effective hydrodynamic size of the migrating species, thereby enhancing their magnetophoretic response beyond what is expected for isolated, hydrated metal ions in solutions. To evaluate this hypothesis, we systematically adjusted the effective metal ion radius $R$ in the simulations to achieve the best agreement with the experimental data. In this approach, the cluster size is treated as a constant, representing a fixed number of ions or molecules aggregated under the influence of the magnetic field. The solid curves in Fig.~\ref{constantcluster}(a,b) represent the results of simulations in which the effective metal ion radius $R$ was optimized to match the experimentally observed separation after 72 hours. These simulations successfully reproduce the key experimental trends, showing enrichment of MnCl$_2$ in the near region, consistent with its paramagnetic nature, and depletion of ZnCl$_2$, in line with its diamagnetic behavior. In addition, Fig.~\ref{constantcluster}(c,d) presents the spatiotemporal evolution of metal ion concentration and the corresponding velocity fields in the computational domain for both MnCl$_2$ and ZnCl$_2$, based on simulations that closely match the experimental data. The simulations indicate that the magnetophoretic velocity increases toward the lower-left corner of the computational domain, where the magnetic field gradient is highest, and that gives rise to the development of a concentration gradient. This velocity is directed toward the magnet (positive) for MnCl$_2$ and away from the magnet (negative) for ZnCl$_2$, consistent with their respective magnetic susceptibilities. These simulations are performed for various initial concentrations and the corresponding values of the constant cluster sizes used for MnCl$_2$ and ZnCl$_2$ in the simulations are summarized in Table~(\ref{constantCluster}). Notably, the averaged cluster size decreases with increasing ion concentration and also larger cluster sizes are predicted for the diamagnetic ZnCl$_2$ metal ions compared to the paramagnetic MnCl$_2$.\par 
\begin{table*}[hthp]

\centering
\caption{\centering The constant cluster sizes used to match the simulations with experiments involving polydisperse porous medium.}
\begin{tabular}{|c|c|c|c|c|c|c|}
\hline
\multirow{2}{*}{c$_0$\text{[mM]}}& \multicolumn{3}{c|}{\textbf{MnCl$_2$}} & \multicolumn{3}{c|}{\textbf{ZnCl$_2$}} \\ \cline{2-7} 
 & \multicolumn{1}{c|}{$R $ [$\mu$m]} & \multicolumn{1}{c|}{$N$} & \multicolumn{1}{c|}{$n$} & \multicolumn{1}{c|}{$R $ [$\mu$m]} & \multicolumn{1}{c|}{$N$} & \multicolumn{1}{c|}{$n$} \\ \hline
1    & 4.00      & 1.6$\times 10^8$ & 1.24$\times 10^{9}$&  38.83  & 1.5$\times 10^{11}$& 1.35$\times 10^{6}$\\ 
10   &  1.07  & 0.31 $\times 10^8$& 6.47$\times 10^{10}$&  9.74  & 0.23$\times 10^{11}$ &  8.57$\times 10^{7}$\\ 
100  &  0.36 & 0.12 $\times 10^8$&  1.68$\times 10^{12}$&  4.29 & 0.20$\times 10^{11}$ &  1.00$\times 10^{9}$ \\ \hline
\end{tabular}
\label{constantCluster}
\end{table*}

Although the simulations successfully reproduce the quasi-steady enrichment or depletion of metal ions after 72 hours, important considerations arise regarding the interpretation of the fitted cluster sizes, particularly their dependence on the initial ion concentration and magnetic susceptibility. Notably, the variation of cluster size with concentration appears counterintuitive. One might expect that, if metal ions aggregate into clusters under a magnetic field, higher initial concentrations would lead to larger average cluster sizes, as observed in previous studies on superparamagnetic nanoparticles where cluster size increases with volume fraction~\cite{promislow1995aggregation}. However, our simulations suggest the opposite: larger effective cluster sizes are required to match experimental observations at lower metal ion concentrations. To investigate this further, we estimated the number of ions per cluster ($N$) based on the fitted hydrodynamic radii, with results summarized in Table~(\ref{constantCluster}). These estimates show that clusters formed at lower concentrations contain more ions per cluster. At the same time, the total number of clusters ($n$) in solution decreases with decreasing concentration, as also shown in Table~(\ref{constantCluster}). Thus, while clusters at lower concentrations are larger, they are fewer in number. In contrast, at higher concentrations, the clusters are smaller but more numerous. A larger number of clusters implies a greater number of magnetically responsive units contributing to the overall magnetophoretic flux, which may compensate for weaker individual magnetic forces on smaller clusters. This analysis underscores that interpreting simulation results requires consideration of both the size and number of clusters. Focusing solely on cluster size may obscure important collective effects that influence magnetophoretic transport behavior. \HM{Komori and Terao~\cite{komori2023cluster} showed through molecular dynamics simulations that, as NaCl concentration increases in an electrolyte system, the fraction of isolated ions decreases while larger aggregates form. In our study, we did not explicitly distinguish isolated ions; however, the simulations indicate that larger clusters appear predominantly at lower concentrations accompanied by a decrease in their population density. This contrast suggests that the applied magnetic field may alter the equilibrium between cluster size and number density distinct from those observed in nonmagnetic systems.} Second, the case of ZnCl$_2$ presents a conceptual challenge that remains as an open question. As a diamagnetic compound, ZnCl$_2$ is not expected to form magnetic clusters under uniform or non-uniform magnetic fields.\par 

While the simulations described above successfully capture the quasi-steady enrichment or depletion of metal ions after 72 hours, they tend to overpredict ion transport during the initial transient phase. This discrepancy likely stems from the simplifying assumption of a constant cluster size throughout the simulation. Prior studies have demonstrated that under an applied magnetic field, clusters of super paramagnetic nano-particles evolve dynamically, with their size varying as a function of time in a power law fashion~\cite{martinez2008kinetic,promislow1995aggregation,fermigier1992structure}. Therefore, to assess the deviation between simulations and experiments during initial transient times, we extend our analysis by characterizing the time-dependent evolution of cluster size. \par

\subsubsection{Time-Dependent Behavior of Metal Ion Clusters}
Previous experiments on superparamagnetic colloidal nanoparticles based on iron oxide have shown that size of the particle cluster varies with time in a power law fashion~\cite{reynolds2015deterministic, faraudo2013understanding}. Therefore, in our simulations, we used a power law model to express metal ion cluster size evolution with time as: $R_t =R_0 + \alpha t^{\beta}$, where $R_t$ represents the cluster size at time $t$, $R_0$ is the initial hydrated metal ion size, $\alpha$ is a scaling factor, and $\beta$ is an empirical exponent governing the cluster growth dynamics. The simulations were then optimized by varying $\alpha$ and $\beta$ values to find the best match with experiments. \par
\begin{figure*}[ht]

    \centering
    \subfloat{{\includegraphics[width=0.6\linewidth]{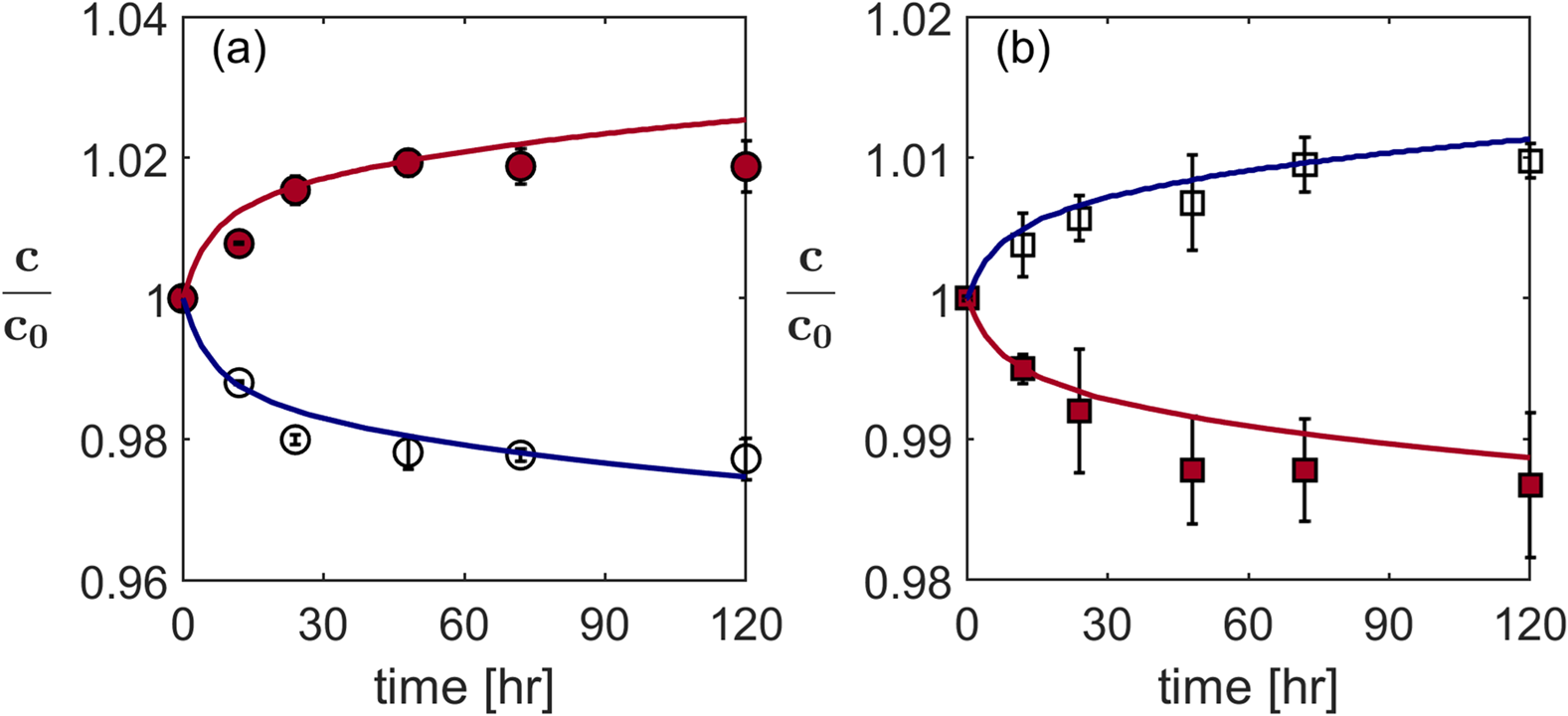}}}\\
 
    \subfloat{{\includegraphics[width=0.7\linewidth]{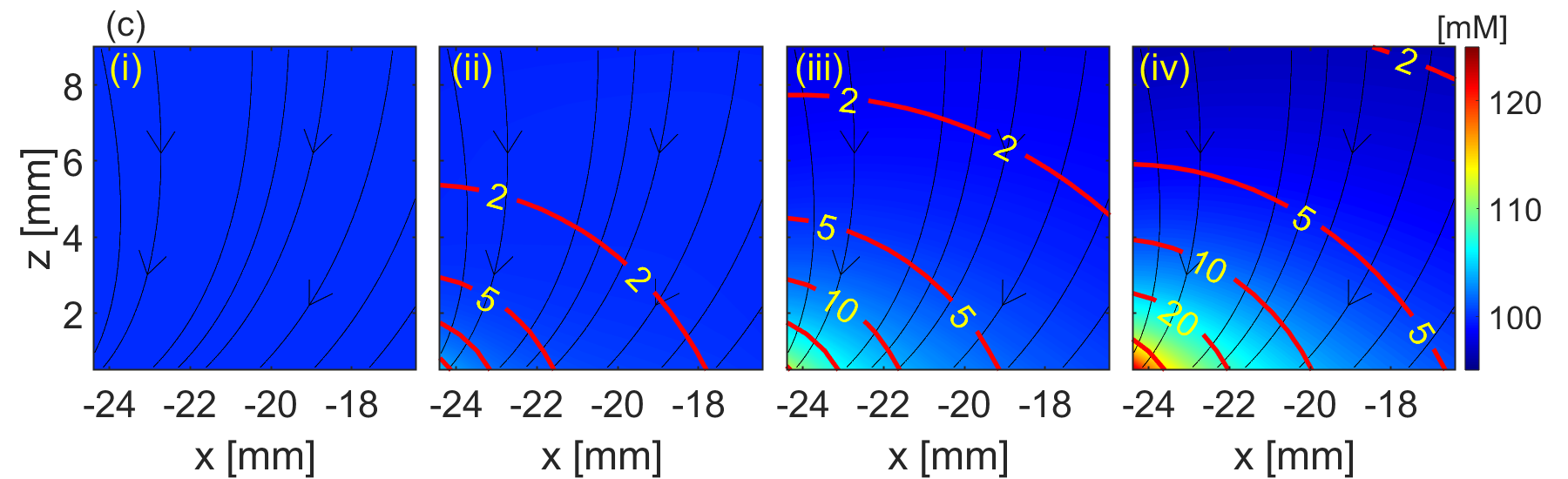}}}\\
    \subfloat{{\includegraphics[width=0.7\linewidth]{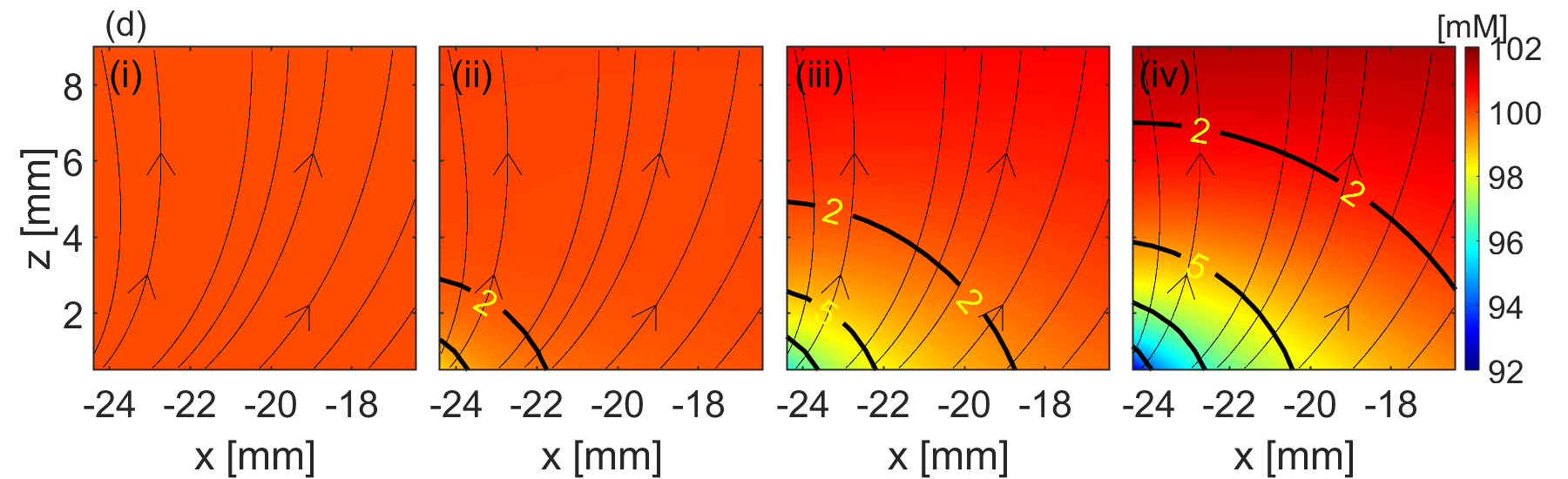}}}\\
    
        \caption{Temporal evolution of metal ion concentration along with the predictions of the model (solid curves) for MnCl$_2$ (a) and ZnCl$2$ (b) at initial concentration c$_\text{0}$ = 100mM. In part (a) and (b), the filled and open symbols denote the concentrations in the near and far halves. The continuous curves denote the model predictions for a dynamically evolving  cluster size. The spatio-temporal evolution of metal ion concentration along with velocity vectors and velocity contours for MnCl$_2$(c) and ZnCl$_2$ (d) for an initial concentration of 100mM after t = 0 (i), t = 1 (ii), t = 12 (iii), and t = 72 (iv) hours. Velocity labels in (c,d) have a unit of nm/s.}
    \label{Variable_cluster}
\end{figure*}

Fig.~(\ref{Variable_cluster}) presents the results of numerical simulations that best match the experimental data for both paramagnetic MnCl$_2$ and diamagnetic ZnCl$_2$ at an initial concentration of 100 mM. A key observation is that the initial transient evolution of metal ion concentration is now more accurately captured compared to previous simulations with a constant cluster size. This analysis was performed for various initial concentrations (see the corresponding concentration plots in Fig.~S1, and Fig.~S2 of the supplementary materials) and the resulting cluster size parameters are summarized in Table~(\ref{Var clu size values}). Several observations can be made here. First, as shown in Table~(\ref{Var clu size values}), the exponent $\beta$ is typically below unity and decreases with increasing initial concentration of metal ions. Different power-law exponents for magnetically induced aggregation have been reported in the literature, particularly for superparamagnetic colloids and nanoparticles. For instance, \citeauthor{promislow1995aggregation} reported exponent values ranging from 0.5 to 0.75 for superparamagnetic polystyrene nanoparticles under varying magnetic fields and initial volume fractions~\cite{promislow1995aggregation}. Similarly, \citeauthor{andreu2011aggregation} found a scaling exponent of approximately 0.64 from Langevin dynamics simulations of superparamagnetic colloids~\cite{andreu2011aggregation}. The power-law exponent is smaller in our study than those reported for super-paramegnetic nanoparticles. This is perhaps not surprising because unlike superparamagnetic colloids, metal ions possess significantly smaller magnetic moments and hydrodynamic radii, which likely contribute to the lower exponents observed in our simulations. Table~(\ref{Var clu size values}) also suggests that the power-law exponent ($\beta$) increases as the initial metal ion concentration decreases. This behavior is consistent with previous findings that showed $\beta$ decreases as the initial concentration of nanoparticles increases in the solution~\cite{promislow1995aggregation, fermigier1992structure}. It has been suggested that at high volume fractions, interparticle interactions are enhanced and the resulting crowding reduces the effective aggregation rate~\cite{promislow1995aggregation, fermigier1992structure}. Finally, we note that similar trends for $\beta$ are observed for diamagnetic metal ions based on ZnCl$_2$. To the best of our knowledge, no magnetically induced cluster formation has been reported for diamagnetic species.\par 
Despite these consistent findings on scaling behavior for paramagnetic metal ions, the corresponding aggregation prefactor ($\alpha$), which defines the initial rate and amplitude of cluster growth, has not been explicitly reported in the literature even for superparamagnetic nanoparticles. Our results indicate that $\alpha$ increases as the initial metal ion concentration decreases. The scarcity of studies involving metal ions highlights the need for further investigation into their aggregation kinetics under magnetic influence.\par 

\begin{table}[hthp]
    \centering
    \caption{Dynamically evolving metal ion cluster sizes obtained from best match with experiments conducted on polydisperse porous medium.}
    \label{Var clu size values}
    \begin{tabular}{|c|c|c|c|c|}
    \hline
    \multirow{2}{*}{c$_0$ \text{[mM]}}  & \multicolumn{2}{c|}{\textbf{MnCl$_2$}} & \multicolumn{2}{c|}{\textbf{ZnCl$_2$}}\\
    \cline{2-5}
      & \( \alpha \) & \( \beta \) & \( \alpha \) & \( \beta \) \\
        \hline
         100 & 0.08 & 0.12 & 0.60 & 0.16 \\
         \hline
         10 & 0.2 & 0.14 & 0.9 & 0.20\\
         \hline
         1 & 0.7 & 0.15 & 2.5 & 0.22 \\
         \hline
    \end{tabular}
    
\end{table}

\subsubsection{Contribution of Kelvin and Paramagnetic Forces}
As noted before, under a non-uniform magnetic field, the magnetic force has two contributions: 
\begin{equation}\label{eqn2}
    \mathbf{F_m} = \frac{4\pi R^3 \Delta\chi_m c}{3\mu_0} (\mathbf{B} \cdot \nabla) \mathbf{B} + \frac{2\pi R^3 \Delta\chi_m \mathbf{B}^2}{3\mu_0} \nabla c 
\end{equation}
The first term on the right-hand side of Eq.~(\ref{eqn2}) denotes the magnetic Kelvin force due to field gradients ($ \mathbf{F}_{\nabla \mathbf{B}}$), while the second term accounts for the contribution from concentration gradients called paramagnetic force ($ \mathbf{F}_{\nabla \mathbf{c}}$). To assess the relative importance of the different magnetic force components in the experiments described in Part (I) of this series, we independently evaluated their contributions using detailed numerical simulations.\par
\begin{figure}[hthp]
    \centering
    \subfloat{{\includegraphics[width=\linewidth]{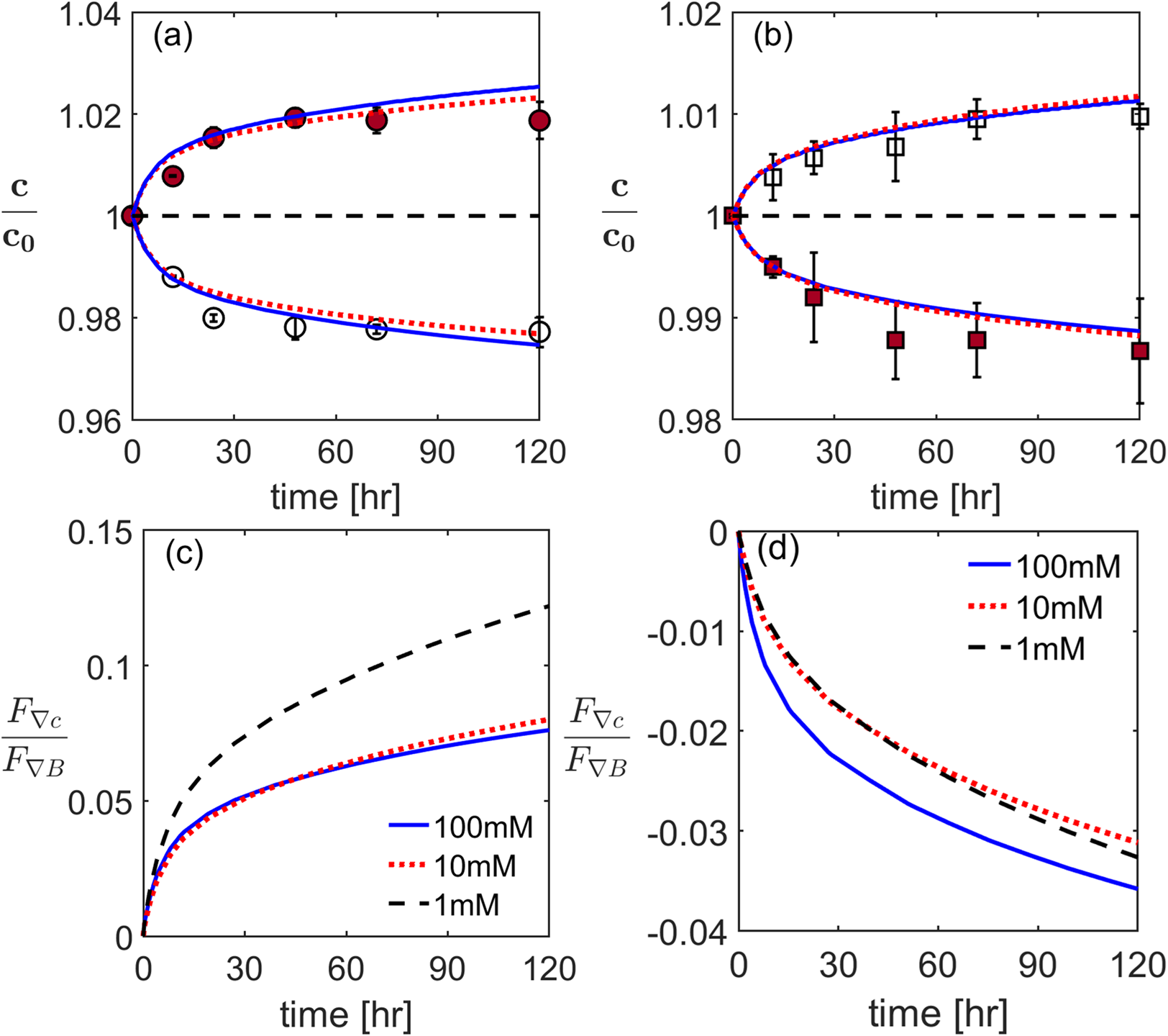}}}\\

    \caption{Temporal evolution of metal ion concentration for (a) MnCl$_2$ and (b) ZnCl$_2$ at c$_0$ = 100 mM. In (a,b) the dashed line, dotted line and the continuous lines correspond to model predictions by including paramagnetic force, Kelvin force and both forces, respectively. The ratio of the paramagnetic force to that of the Kelvin force as a function of time for MnCl$_2$ (c) and ZnCl$_2$ (d) at various initial concentrations. }
    \label{Flux Contribution}
\end{figure}

 
To this end, we first performed simulations by setting the first term $\mathbf{F}_{\nabla \mathbf{B}} = 0$, and applied a concentration perturbation at two locations within the computational domain while conserving total mass (see Fig.~S3 in the supplementary materials). The resulting concentration evolution is plotted as the horizontal black dashed line in Fig.~\ref{Flux Contribution}(a). It is clear that with no magnetic field gradients, no significant separation occurs, and the overall concentration of species in the medium becomes homogeneous within 10 minutes due to diffusion of species. Subsequently, we performed simulations by taking into account the contribution of the field gradient force only, using the parameters summarized in Table~(\ref{Var clu size values}). As shown in Fig.~\ref{Flux Contribution}(a), the dotted curve represents transport driven solely by the Kelvin force, while the thick solid curve includes both the Kelvin and the paramagnetic forces. For the paramagnetic ion, the combined influence of both forces led to further enhancement of separation, illustrating a synergistic effect between the two magnetic forces. \par 
For the diamagnetic ZnCl$_2$, we observe similar trends in that the paramagnetic force alone is insufficient to induce any significant magnetophoretic motion. Interestingly, when simulations include only the Kelvin force (arising from concentration gradients in a non-uniform magnetic field), the predicted extent of magnetophoresis slightly overestimates the experimental results. However, when the paramagnetic force is also included, the predicted separation aligns more closely with the observed data. This suggests that, as concentration gradients evolve within the system, the Kelvin and paramagnetic forces may act in opposition for the diamagnetic metal ions, leading to a subtle competition that influences the net transport behavior. To further evaluate the relative importance of these two forces, the ratio of these two forces could be defined as: 
\begin{equation}\label{Flux ratio formula}
    \frac{F_{\nabla c}}{F_{\nabla B}} = \left | \frac{\mathbf{B}^2}{2c} \frac{\nabla c}{\mathbf{B}\cdot\nabla\mathbf{B}} \right |
\end{equation}

Fig.~\ref{Flux Contribution}(c,d) presents the time evolution of the domain-averaged ratio between the paramagnetic and Kelvin forces for both paramagnetic and diamagnetic metal ions. As expected, the paramagnetic force is significantly smaller than the Kelvin force throughout the simulations for both ion types. However, its relative contribution increases with time, particularly in low concentration regimes. For MnCl$_2$, the paramagnetic force remains modest: contributing less than 8\% of the total force at 100 mM, under 10\% at 10 mM, and up to 20\% at 1mM. These results indicate that while the paramagnetic force is secondary in magnitude, it plays a non-negligible role in modulating magnetophoretic behavior, especially under dilute conditions where concentration gradients become more pronounced. For the diamagnetic metal ion (ZnCl$_2$), the ratio of paramagnetic to Kelvin force is negative, indicating that the two forces act in opposing directions. Similar to the MnCl$_2$ case, the magnitude of the paramagnetic force increases over time but remains small relative to the dominant Kelvin force. \HM{The spatio-temporal evolution of concentration and velocity fields are shows in Fig.~S3 and Fig.~S4 of the supplementary materials. In addition, the analysis suggests that the paramagnetic force likely had a negligible influence in our previous model~\cite{Rassolov2024magnetophoresis} and would not have affected its predictions. }\par 

\subsubsection{Effect of Porous Media}
\begin{figure}[hthp]
    \subfloat{{\includegraphics[width=\linewidth]{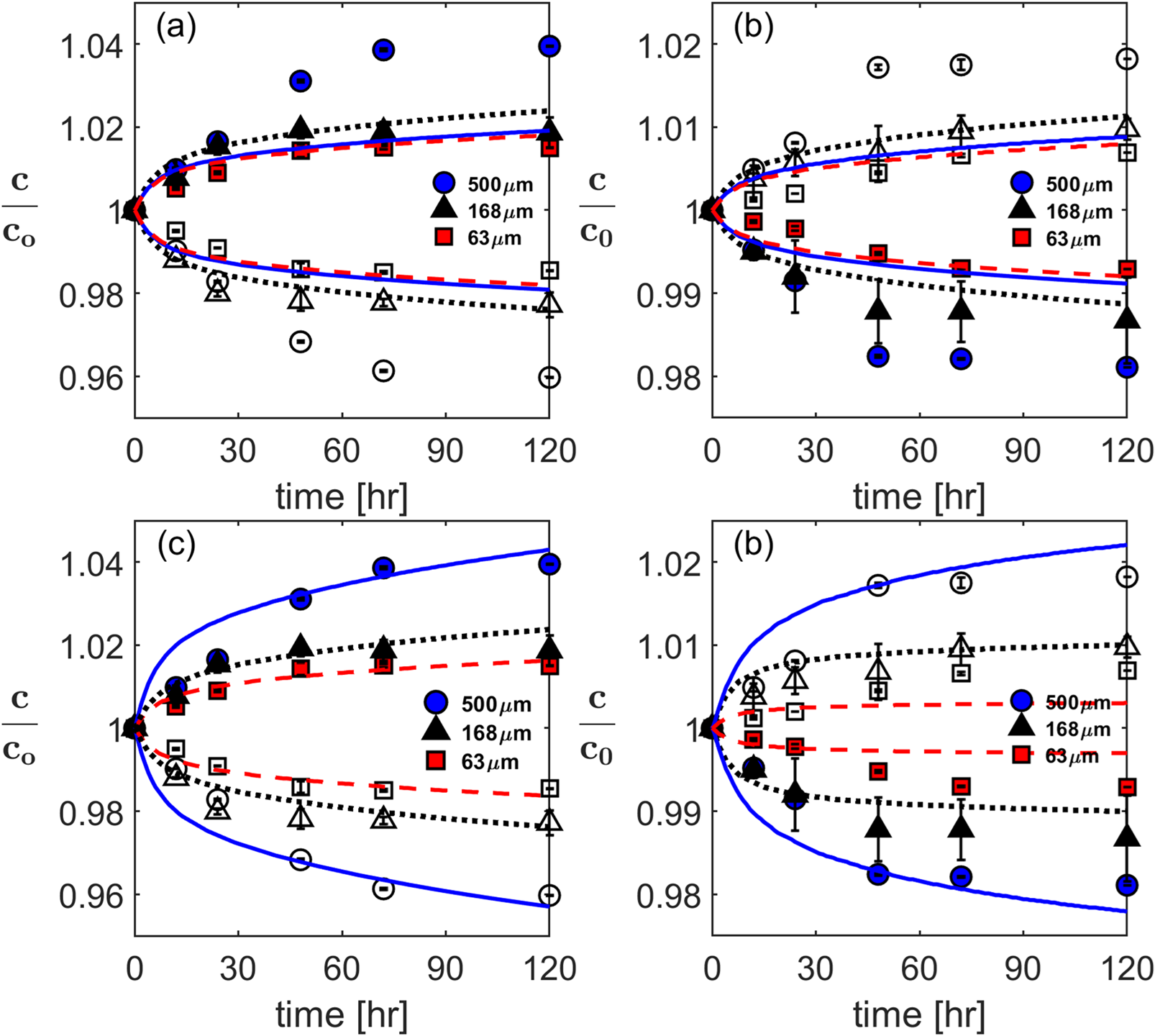}}}\\
   
    \caption{Temporal evolution of concentration for MnCl$_2$ (a) and ZnCl$_2$ (b) in various porous media along with the corresponding predictions of Stokes' Formulation. In part (c,d), the corresponding calculations are based on Brinkman's formulation. In (a-d), the dashed, dotted and continuous lines denote simulations for porous media with 63 $\mu$m, 168 $\mu$m and 500 $\mu$m particle size respectively.}
    \label{Stokes and brinkman formulation}
\end{figure}

Up to this point, we have employed the Stokes formulation to model the magnetophoresis of metal ions within a porous medium characterized by a polydisperse particle size distribution. In Part (I) of this study~\cite{Nwachuwku2025}, we reported experimental observations of magnetophoretic transport of metal ions through porous media composed of particles with varying sizes. The results demonstrated that larger particle sizes enhance the magnetophoretic response, indicating a strong dependence of ion transport on the microstructure of the porous medium. In the present simulations, the effect of the porous matrix is incorporated via its influence on Stokes drag through the effective diffusivity $D_e$. Using experimentally measured porosity values, the numerical simulations based on Stokes' formulations were performed for various porous media and results are shown in Fig.~\ref{Stokes and brinkman formulation}(a,b) as solid curves. In all simulations, we assumed a fixed temporal cluster size evolution for each porous medium, as summarized in Table~\ref{Var clu size values}, meaning that porosity is the only variable distinguishing between the cases. Under these assumptions, the simulations predict the highest magnetophoretic transport for the polydisperse medium (with an average particle size of 168 $\mu$m), followed by the 500 $\mu$m medium, and the lowest transport for the medium composed of 63 $\mu$m particles. These predictions deviate from the experimental trend, which showed that larger particle sizes correlate with increased magnetophoretic transport. The discrepancy suggests that porosity alone may not fully capture the influence of porous media structure on transport behavior, and additional factors such as permeability of the porous media may need to be incorporated.\par 
To address this discrepancy, we employed a macroscopic framework that incorporates a critical porous media property—permeability—which is not accounted for in the classical Stokes formulation. Fluid flow in porous structures is commonly described by Darcy’s law~\cite{HUBBERT01031957}, which introduces permeability as a governing parameter. However, previous studies have shown that Darcy’s law becomes inadequate near boundaries, where it fails to capture viscous shear effects~\cite{almalki2016investigations}. To overcome this limitation, we adopted the Brinkman equation, which augments Darcy’s law by incorporating viscous dissipation within the porous medium~\cite{brinkman1949calculation}. This model effectively bridges the behavior of slow flow in porous media and free-fluid Stokes flow. The first step in implementing the Brinkman model is to estimate the permeability of the porous medium. As an initial approach, we employed the Carman-Kozeny relation, which relates permeability to porosity, sphericity, and particle size~\cite{rehman2024comprehensive}. The resulting permeability estimates are presented in Table~\ref{permeability values}. However, simulations using these Carman-Kozeny-derived values did not replicate the experimental trends (see Fig.~S7 in the supplementary materials). In fact, in those simulations a similar trend to those reported by Stokes' formulations is predicted. This mismatch is likely due to overestimation of the permeability values via Carman-Kozney relation. The overestimated permeability values led to negligible damping terms ($\lambda$) in the Brinkman equation, effectively reducing the model to the Stokes problem. \par 
The Carman-Kozney model assumes idealized, uniform pore structures and does not adequately account for the geometric complexity and tortuosity of real porous media. Previous studies have reported permeability values for silica-based porous media in the range of 10$^{-15}$–10$^{-14}$m$^2~$~\cite{kaminska2017dynamic}, which are significantly lower than those predicted by this model. To obtain more accurate estimates of permeabilities, we conducted experiments to directly measure the permeability of the porous media (see the Supplementary materials for experimental details). The measured values, listed in Table~\ref{permeability values}, confirm that the Carman–Kozeny model substantially overestimates permeability in our system. Incorporating these experimentally determined values into the Brinkman model, we carried out numerical simulations of magnetophoresis. As shown in Fig.~\ref{Stokes and brinkman formulation}(c,d), the updated simulations successfully reproduce the experimental trends, including the enhanced magnetophoresis observed in porous media composed of larger particles. \HM{The corresponding spatio-temporal evolution of concentration and velocity profiles for Brinkmann model are presented in Fig.~S5 and Fig.~S6 of the supplementary materials. } These results clearly demonstrate the importance of employing a more physically realistic Brinkman formulation to accurately capture the influence of porous media structure and predict experimental observations.\par 
    
\begin{table*}[hthp]
    \centering
    \caption{A summary of porous media properties that are used in this study. }
    \label{permeability values}
    \begin{tabular}{|c|c|c|c|c|}
    \hline
    \multirow{2}{*}{Porous media} & porosity  & Average size  & Carman-Kozeny   & Experiments of this work \\
     & $\varepsilon$ [-] & [$\mu$m] & $\kappa \times 10^{12}  [m^2] $ & $\kappa \times 10^{12}  [m^2] $ \\
    \hline
        Poly-dispersed  & 0.61 & 168 &  231.22 & 7.20 $\pm$ 1.54\\
        \hline
        63 microns & 0.74 & 63 & 132.18 & 2.74 $\pm$ 0.04 \\
        \hline
        500 microns & 0.71  & 500 & 5910.80 & 39.58 $\pm$ 1.51 \\
        \hline
    \end{tabular}
    
\end{table*}
\subsection{Magnetophoresis of Binary Metal ion Mixtures}
In binary electrolyte solutions containing similarly charged cations, classical electrostatic theory predicts mutual repulsion and spatial separation due to like-charge interactions. However, molecular dynamics simulations and experiments have shown that even like-charged ions can exhibit correlated behavior and form transient, dynamically fluctuating ion clusters, particularly at moderate to high concentrations~\cite{chen2007quantitative,bian2011ion,hassan2008computer}. Our experimental observations reported in part (I) of this series lend support to this prediction because they reveal that both Mn$^{+2}$ and Zn$^{+2}$ ions, despite having different magnetic susceptibilities (paramagnetic vs. diamagnetic), exhibit coordinated transport toward the same region of the porous medium. This behavior suggests the existence of like-charge ion–ion correlations or clustered assemblies that defy classical expectations of purely repulsive interactions and may instead facilitate collective transport. Based on these findings, we hypothesize that intermolecular interactions in such binary mixtures facilitate the formation of ionic clusters that may be composed of both paramagnetic and diamagnetic species. These mixed clusters could exhibit an effective magnetic susceptibility different from that of the isolated ions, thereby modifying the overall magnetic response of the system under external fields. To evaluate this hypothesis, we define the molar magnetic susceptibility of the clusters in the binary mixture $\chi_{mm}$ as: 
\begin{equation}
\label{mixture_magn_velocity}
    \chi_{mm}  = \chi_{mp} y_p + \chi_{md} y_d, 
\end{equation}
where $\chi_{mp}$, $\chi_{md}$, $y_p$ and $y_d$ are the molar magnetic susceptibility of the paramagnetic, diamagnetic and mole fraction of paramagnetic and diamagnetic metal ions in the mixture, respectively. While it is not possible to determine the exact composition of individual ion clusters in the binary mixture, we assume that the mole fraction of each metal ion within a cluster reflects its corresponding bulk mole fraction in solution. This approximation allows us to estimate an effective magnetic susceptibility for the mixed clusters based on the weighted contributions of their constituent ions. Using this effective susceptibility, we performed numerical simulations to assess the magnetophoretic transport of each species within the mixture. In these simulations, the cluster size was treated as a dynamic parameter and systematically varied to identify the value that best reproduces the experimental transport profiles.\par
\begin{figure*}[hthp]
    \centering
    \subfloat{{\includegraphics[width=0.9\linewidth]{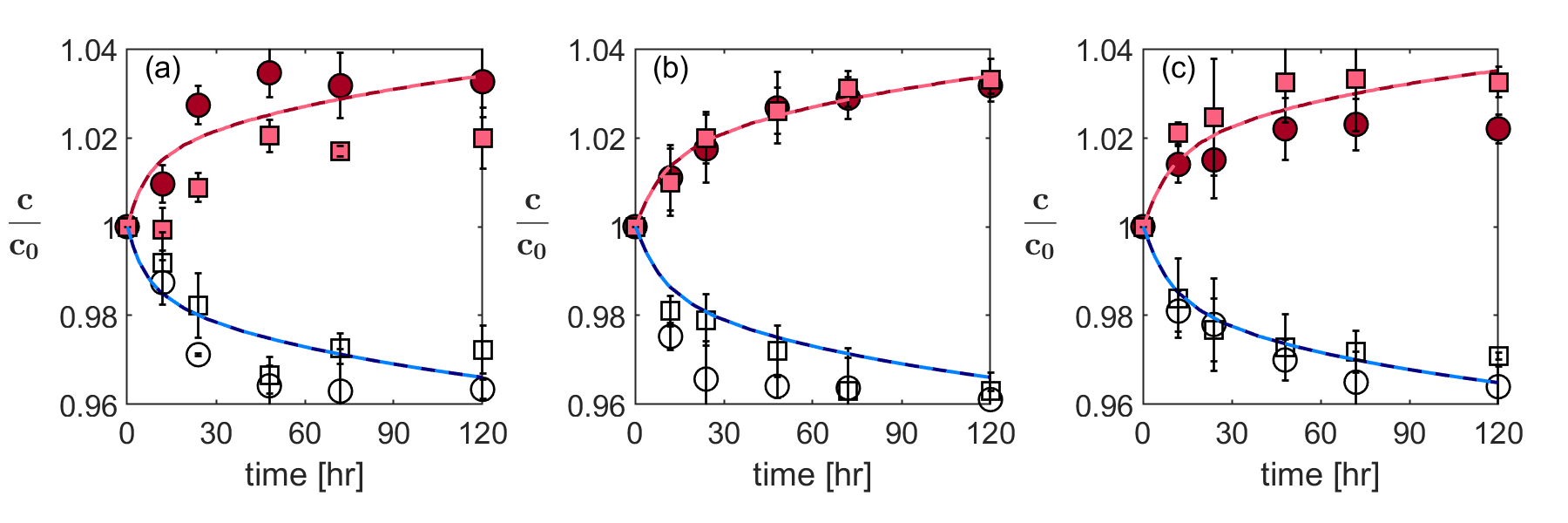}}}
 
    \caption{Temporal evolution of metal ion concentration in binary mixtures for (a) 100mM/100mM MnCl$_2$/ZnCl$_2$, (b) 100mM/10mM MnCl$_2$/ZnCl$_2$, and (c) 10mM/100mM MnCl$_2$/ZnCl$_2$ mixtures. The filled and empty symbols correspond to the near and far sides. In addition, square and circle symbols correspond to ZnCl$_2$ and MnCl$_2$, respectively. The continuous and dashed line correspond to predictions of the model for MnCl$_2$ and ZnCl$_2$, respectively. }
    \label{Mixtures}
\end{figure*}

Fig.~(\ref{Mixtures}) presents the temporal evolution of paramagnetic MnCl$_2$ and diamagnetic ZnCl$_2$ metal ion concentrations in binary mixtures, along with predictions from numerical simulations. At equimolar concentrations, experimental results indicate that MnCl$_2$ is more strongly enriched near the magnet surface than ZnCl$_2$, suggesting a differential response to the magnetic field. When the ZnCl$_2$ concentration is reduced, the degree of MnCl$_2$ enrichment remains largely unchanged, while ZnCl$_2$ shows increased accumulation, eventually reaching a level comparable to that of MnCl$_2$. Upon further reduction of MnCl$_2$ concentration, ZnCl$_2$ enrichment slightly exceeds that of MnCl$_2$. In contrast, the simulations predict similar enrichment behavior for both metal ions across all concentration ratios. This discrepancy may arise from a key model assumption, namely, that the mole fraction of each metal ion within the cluster mirrors its bulk mole fraction in solution. Despite these differences, the simulations reasonably reproduce the overall trends observed in the experiments. The remaining discrepancies likely reflect limitations in the simplified cluster composition assumption and highlight the need for more detailed understanding of ion-specific interactions and clustering behavior in mixed systems.\par
Finally, we revisit the cluster size values for the binary mixtures, as summarized in Table~\ref{Mixturecluster}. For mixtures primarily composed of MnCl$_2$, the extracted cluster sizes remain comparable to those obtained for MnCl$_2$ alone (Table~\ref{Var clu size values}), indicating that the presence of ZnCl$_2$ at lower concentrations does not significantly influence the clustering behavior of Mn$^{+2}$. In these cases, the cluster size appears largely insensitive to ZnCl$_2$ concentration. In contrast, for mixtures dominated by ZnCl$_2$, a slightly larger cluster size is required to reproduce the observed enrichment, relative to other binary compositions. Interestingly, while the rate of temporal cluster growth, characterized by the exponent $\beta$, remains similar across all binary mixtures, the growth rate in ZnCl$_2$-dominant binary mixtures is noticeably slower than that observed in the single-ion ZnCl$_2$ system (Table~\ref{Var clu size values}). This observation suggests that the presence of MnCl$_2$ may alter the dynamics of field-induced clustering in ZnCl$_2$-rich systems, potentially by modifying local ion–ion correlations, hydration structure, or the effective magnetic response of the mixture. \HM{Recent theoretical work by Dinpajooh et al.~\cite{dinpajooh2025magnetic} showed that when paramagnetic ions form localized nanoscale domains, their collective magnetic dipole moments can yield dipole-dipole interactions comparable in magnitude to electrostatic and van der Waals forces at sub-nanometer separations. Such interactions, which depend on the field orientation, can influence aggregation and correlated motion among paramagnetic clusters. While our continuum-based formulation does not explicitly resolve such nanoscale magnetic potentials, the physical mechanism identified in that study may provide a microstructural basis for the observed field-dependent correlations and supports our interpretation that magnetically induced clustering may contribute to the collective transport of paramagnetic and diamagnetic species in complex mixtures.
}

    

\begin{table}[hthp!]
    \centering
    \caption{Dynamically evolving metal ion cluster sizes for binary mixtures obtained from best match with experiments on porous media with a particle size of 500 $\mu$m.}
    \label{Mixturecluster}
    \begin{tabular}{|c|c|c|c|}
    \hline
     \multicolumn{2}{|c|}{c$_0$ \text{[mM]}} &\multirow{2}{*} {\( \alpha \)} & \multirow{2}{*}{\( \beta \)}\\
    \cline{1-2}
   \textbf{MnCl$_2$} & \textbf{ZnCl$_2$} & &\\
        \hline
        10 & 100 & 0.25   & 0.15 \\
         \hline
       100  & 10 & 0.07 &  0.15 \\
         \hline
       100  & 100 & 0.07 & 0.15 \\
         \hline
    \end{tabular}
    
\end{table}

\section*{Conclusion and Summary}
In summary, we have developed a two-dimensional multiphysics numerical model to investigate the magnetophoresis of paramagnetic and diamagnetic metal ions in porous media under the influence of a static magnetic field generated by a permanent magnet. Our study examines both individual metal ions and their binary mixtures across a concentration range of 1 mM to 100 mM. Consistent with recent experimental findings~\cite{Nwachuwku2025}, the simulations show that paramagnetic MnCl$_2$ becomes enriched near the magnet surface, while diamagnetic ZnCl$_2$ is depleted in that region.\par 

The simulations predict negligible magnetophoresis for isolated metal ions with hydrodynamic sizes comparable to their hydrated radii. Agreement with experimental trends is only achieved when field-induced cluster formation is considered, suggesting that metal ions form micron-sized clusters under magnetic fields. Furthermore, the cluster size evolves over time and follows a power-law dependence $R\sim t^n$ consistent with previous studies on super-paramagnetic nanoparticles~\cite{ezzaier2017two, heinrich2015effects, amira2012dynamic}. Our analysis also reveals that the Stokes-based formulation is insufficient to capture the role of porous media properties in magnetophoresis. In contrast, simulations based on the more physically realistic Brinkman model, which incorporates medium permeability, successfully reproduces experimental trends for both paramagnetic and diamagnetic ions across porous media with different particle sizes. Finally, simulations of binary metal ion mixtures reveal that interactions between paramagnetic and diamagnetic clusters significantly influence their transport. Specifically, the presence of diamagnetic clusters reduces the magnetophoretic motion of paramagnetic clusters and vice versa, indicating a hydrodynamic coupling that modulates migration in mixed-ion systems.\par 
Together, these findings highlight the critical role of field-induced metal ion cluster formation, porous media structure, and interspecies interactions in governing magnetophoresis, and provide a predictive framework for understanding and designing magnetically driven transport in complex ionic systems.\par

\section*{Acknowledgements}
A portion of this work was performed at the National High Magnetic Field Laboratory, which is supported by the National Science Foundation Cooperative Agreement No. DMR-1644779 and the state of Florida. This work was supported by the Center for Rare Earths, Critical Minerals, and Industrial Byproducts, through funding provided by the State of Florida. HM acknowledges the support from National Science Foundation through award CET 2343151.

\bibliography{Arxiv}






\clearpage
\onecolumngrid

\setcounter{page}{1}
\setcounter{section}{0}
\setcounter{table}{0}
\setcounter{figure}{0}
\setcounter{equation}{0}

\renewcommand{\thepage}{P\arabic{page}}
\renewcommand{\thesection}{S\arabic{section}}
\renewcommand{\thetable}{S\arabic{table}}
\renewcommand{\thefigure}{S\arabic{figure}}
\renewcommand{\theequation}{E\arabic{equation}}

\renewcommand{\familydefault}{\sfdefault}

\setstretch{1.3}


\begin{center}
    {\LARGE \textbf{Supplementary Materials: Magnetically Assisted Separation of Weakly Magnetic Metal Ions in Porous Media.}}\\[6pt]
    {\large \textbf{Part 2: Numerical Simulations}}\\[12pt]
    {\normalsize Muhammad Jamiu Garba\textsuperscript{1,2}, Alwell Nwachukwu\textsuperscript{1,2}, Jamel Ali\textsuperscript{1,2}, \\Theo Siegrist\textsuperscript{1,2}, Munir Humayun\textsuperscript{2,3}, Hadi Mohammadigoushki\textsuperscript{1,2,*}}\\[8pt]
    {\small \textsuperscript{1}Department of Chemical and Biomedical Engineering, FAMU-FSU College of Engineering, Tallahassee, FL\\
    \textsuperscript{2}National High Magnetic Field Laboratory, Tallahassee, FL\\
    \textsuperscript{3}Department of Earth, Ocean and Atmospheric Science, Florida State University, Tallahassee, FL\\
    *Corresponding author: hadi.moham@eng.famu.fsu.edu}\\[12pt]
    {\small \today}
\end{center}

\vspace{1em}
\vspace{2em}

\section*{Additional Simulation Results}
\begin{figure}[H]
    \centering
    \subfloat{{\includegraphics[width=\linewidth]{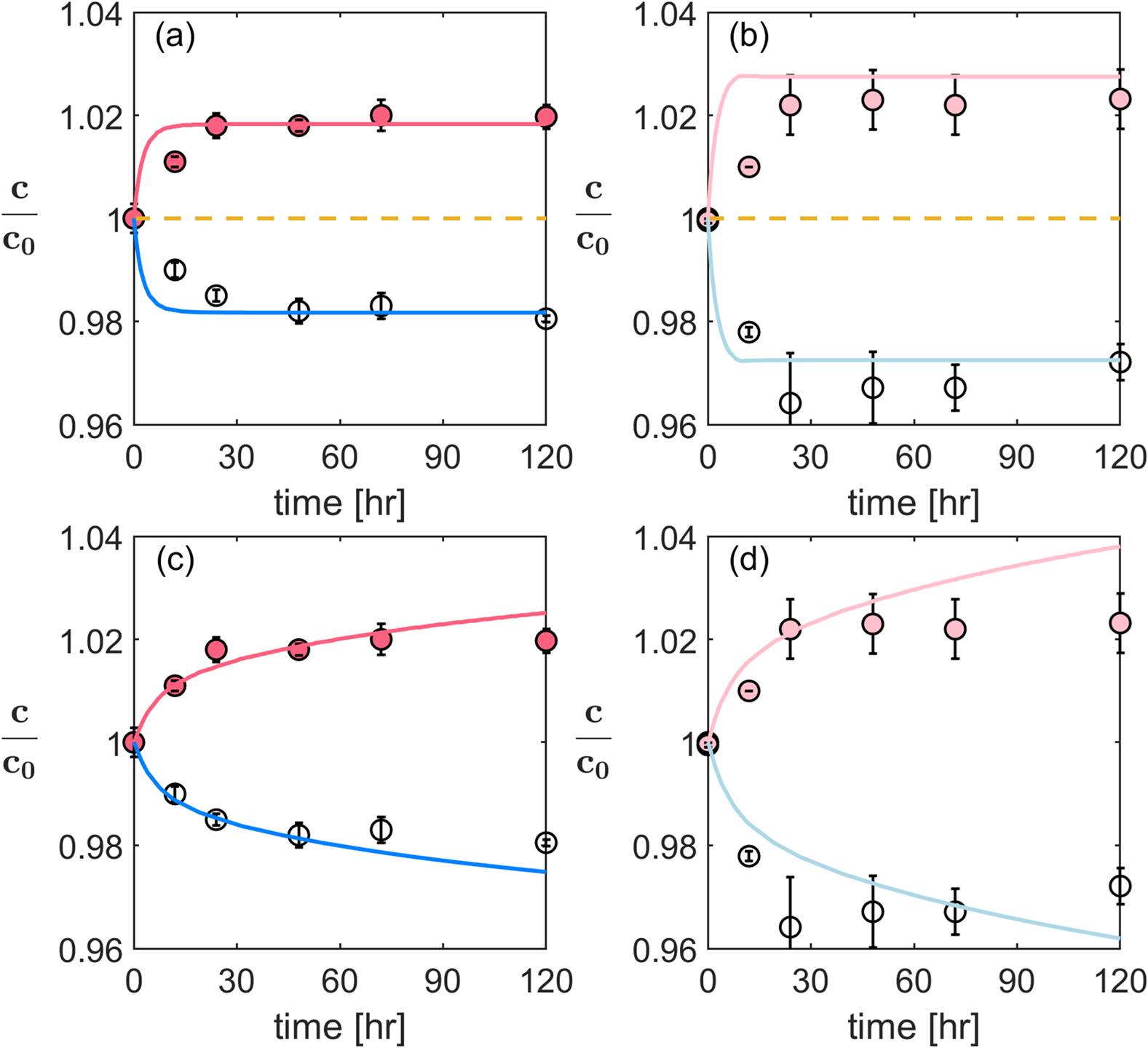}}}\\
   
    \caption{Temporal evolution of MnCl$_2$ as reported in experiments for initial concentration of 10mM (a,c) and 1mM (b,d), and numerical simulations (curves) with a constant cluster size (a,b), and a dynamically evolving cluster size (c,d). Closed symbols correspond to near section and empty symbols are for the far section of the domain.}
    \label{fig:MnCl2-10-1mM}
\end{figure}

\begin{figure}[H]
    \centering
    \subfloat{{\includegraphics[width=\linewidth]{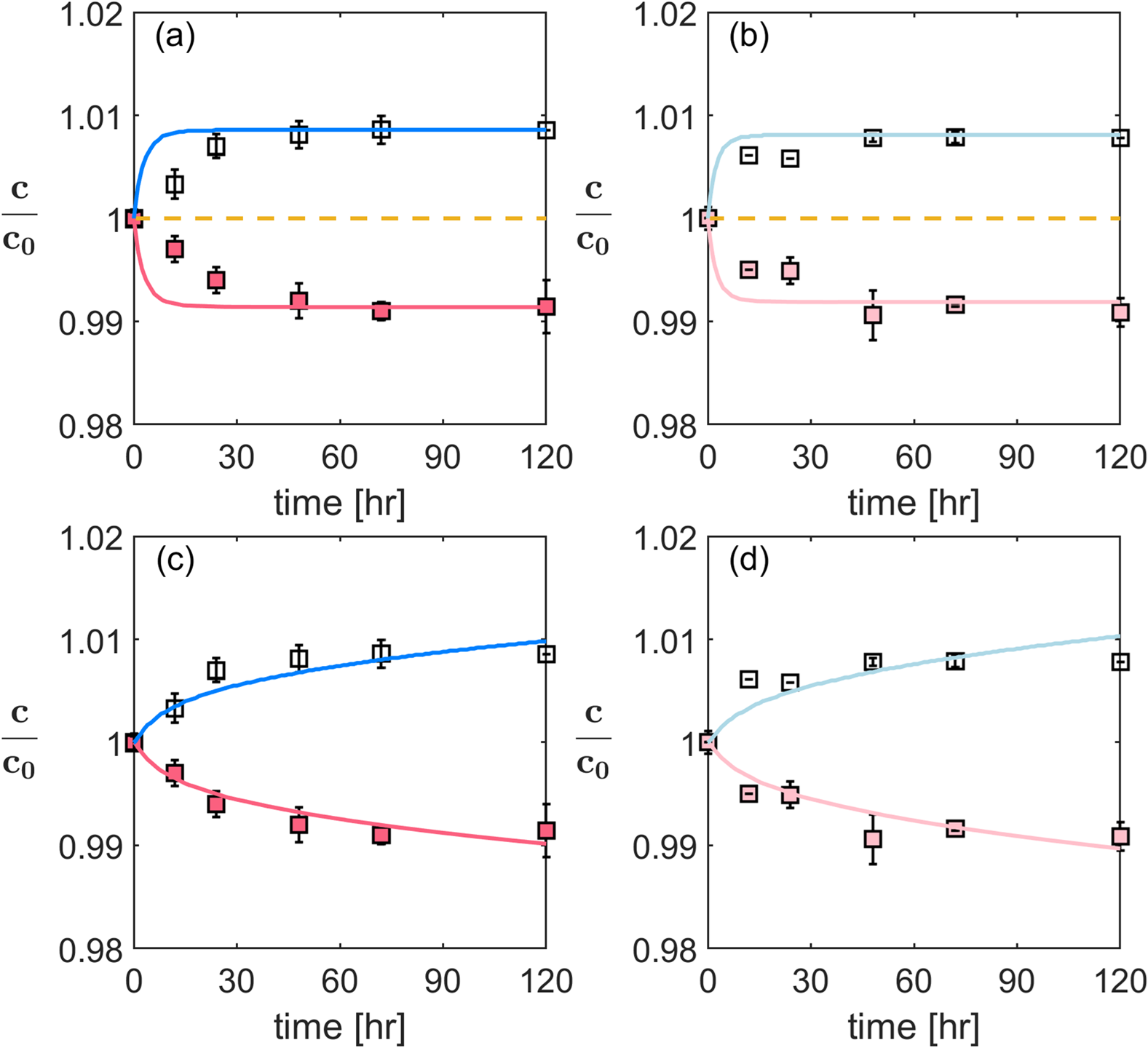}}}\\

    \caption{Temporal evolution of ZnCl$_2$ as reported in experiments for initial concentration of 10mM (a,c) and 1mM (b,d), and numerical simulations (curves) with a constant cluster size (a,b), and a dynamically evolving cluster size (c,d). Closed symbols correspond to near section and empty symbols are for the far section of the domain.}
    \label{fig:ZnCl2-10-1mM}
\end{figure}


\begin{figure}[H]
    \centering
    \subfloat{\includegraphics[width=\linewidth]{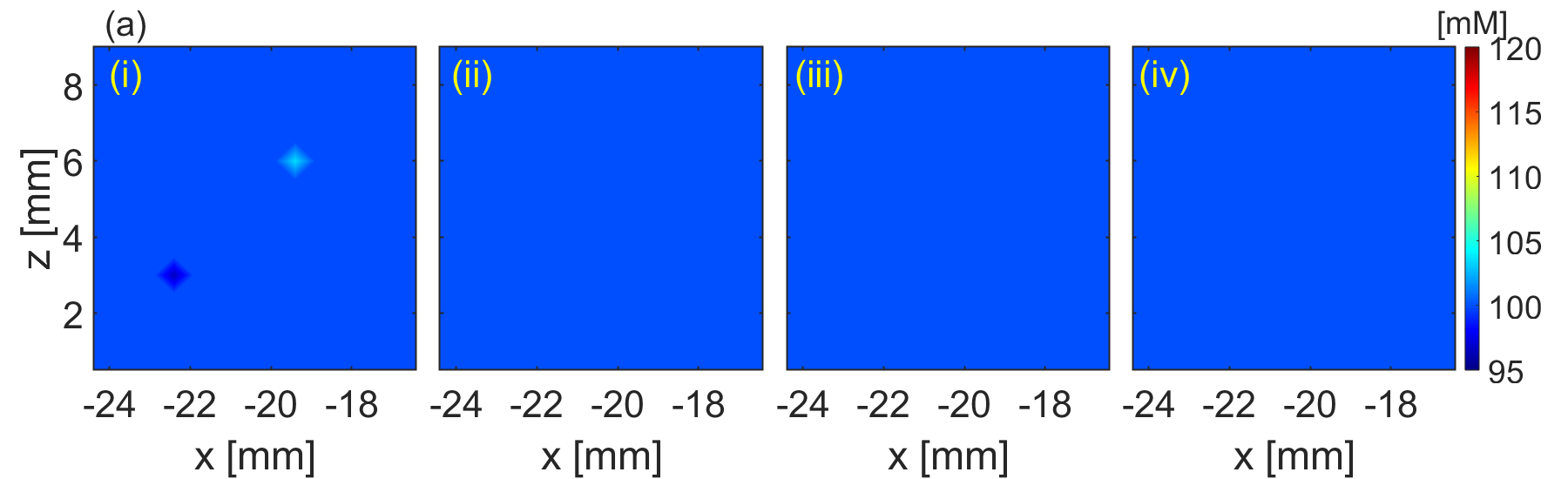}} \\
    \subfloat{\includegraphics[width=\linewidth]{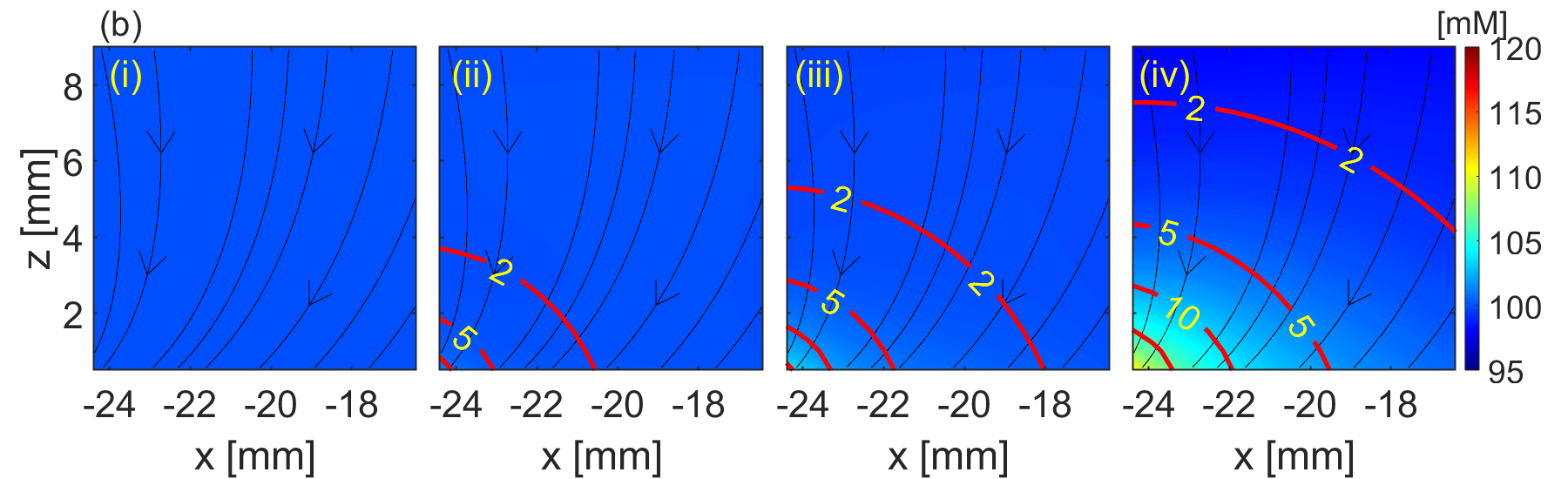}} \\
    \subfloat{\includegraphics[width=\linewidth]{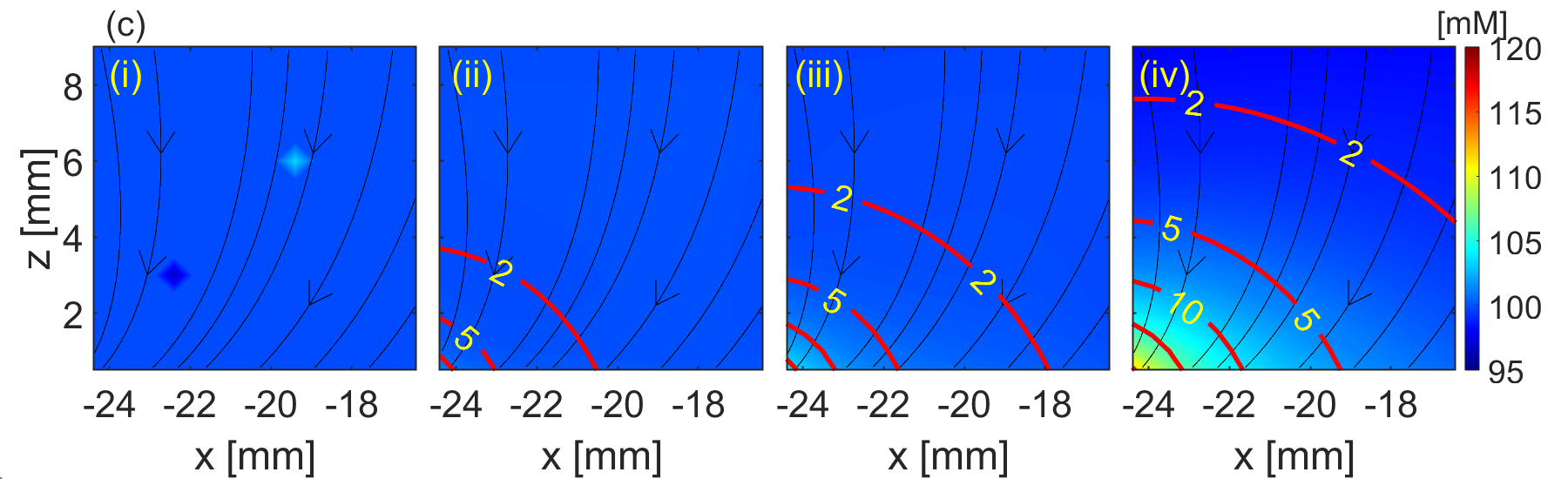}} \\
    \caption{Spatio-temporal evolution of MnCl$_2$ concentration under different magnetic force contributions for an initial concentration of 100~mM. (a) Paramagnetic force only ($F_{\nabla c}$), (b) Kelvin force only ($F_{\nabla B}$), and (c) combined paramagnetic and Kelvin forces. Each subpanel shows heatmaps of concentration along with velocity vectors and contours at four time points: $t = 0$ (i), 0.1 (ii), 1 (iii), and 12 (iv) hours. The color gradients represent ion enrichment toward the magnet. Velocities are shown in nm/s.}
    \label{fig:MnCl2-magnetic-forces}
\end{figure}

\begin{figure}[H]
    \centering
    \subfloat{\includegraphics[width=\linewidth]{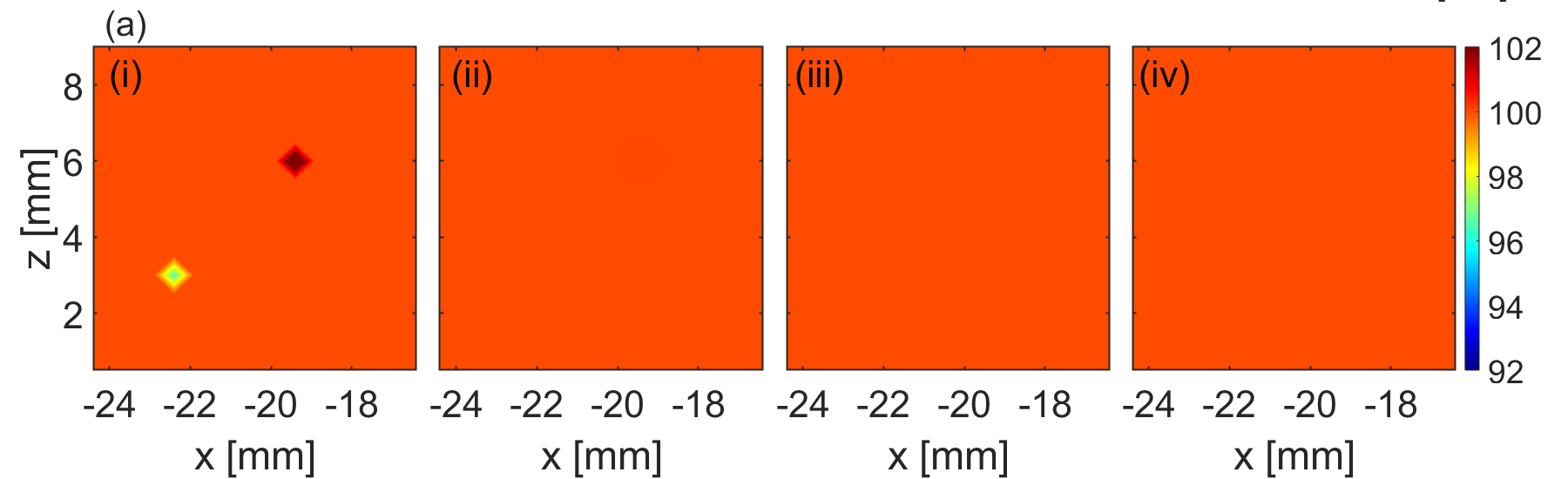}} \\
    \subfloat{\includegraphics[width=\linewidth]{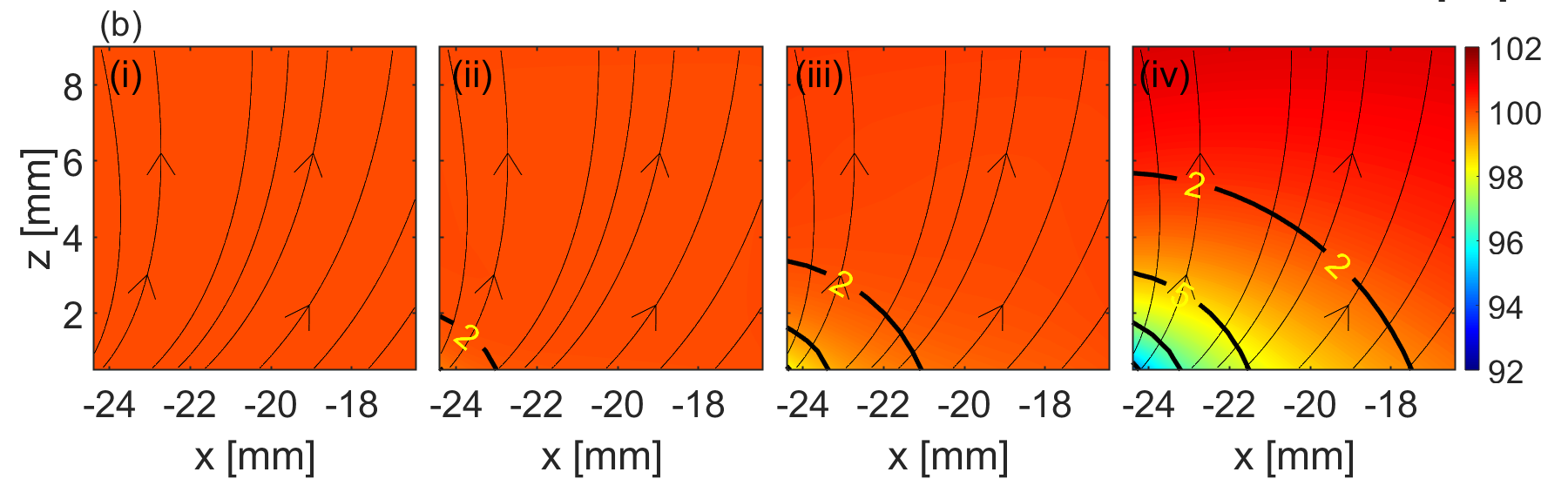}} \\
    \subfloat{\includegraphics[width=\linewidth]{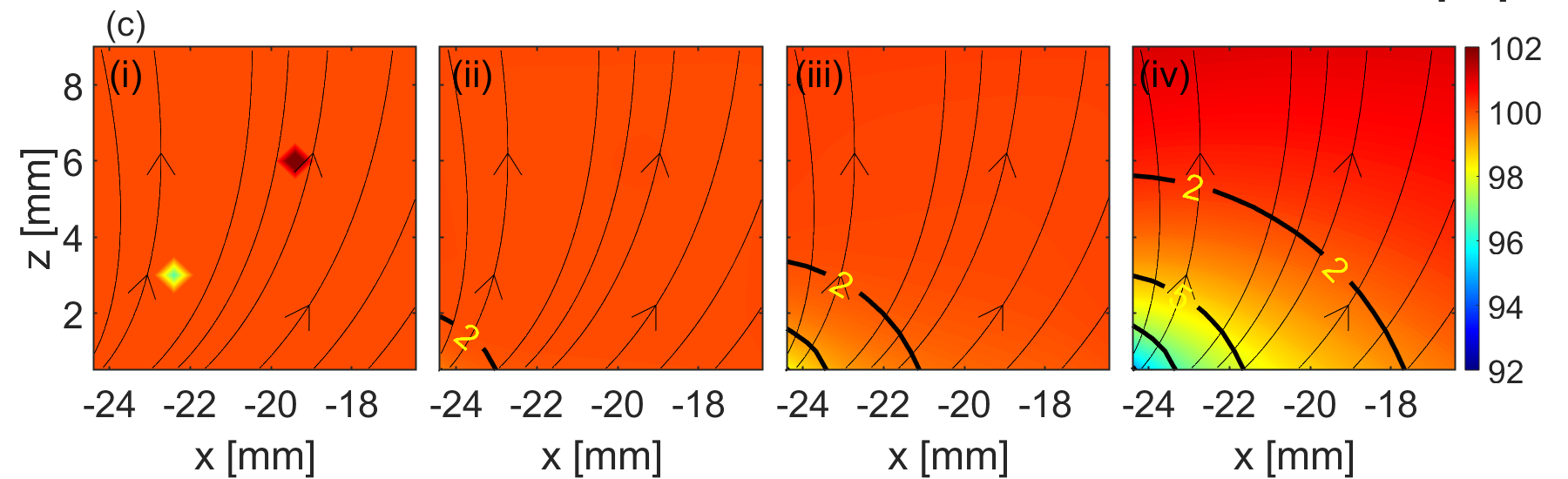}} \\
    \caption{Spatio-temporal evolution of ZnCl$_2$ concentration under different magnetic force contributions for an initial concentration of 100~mM. (a) Paramagnetic force only ($F_{\nabla c}$), (b) Kelvin force only ($F_{\nabla B}$), and (c) combined paramagnetic and Kelvin forces. Each subpanel shows heatmaps of concentration along with velocity vectors and contours at four time points: $t = 0$ (i), 0.1 (ii), 1 (iii), and 12 (iv) hours. The color gradients represent ion depletion toward the magnet. Velocities are shown in nm/s.}
    \label{fig:ZnCl2-magnetic-forces}
\end{figure}

\begin{figure}[H]
    \centering
    \subfloat{\includegraphics[width=\linewidth]{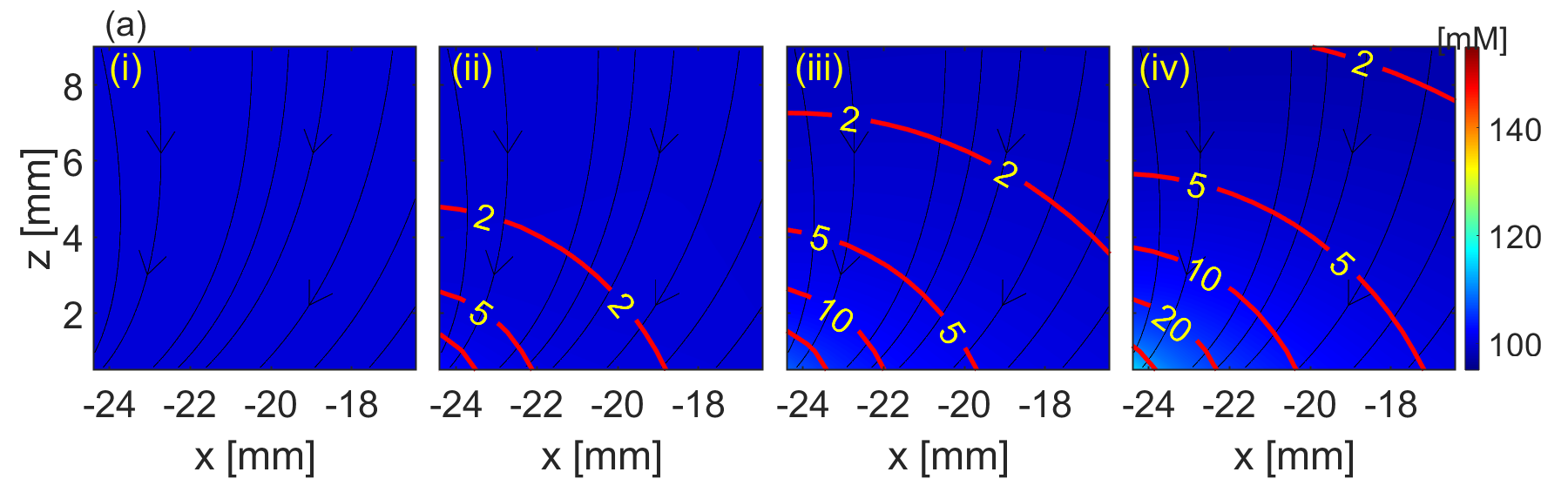}} \\
    \subfloat{\includegraphics[width=\linewidth]{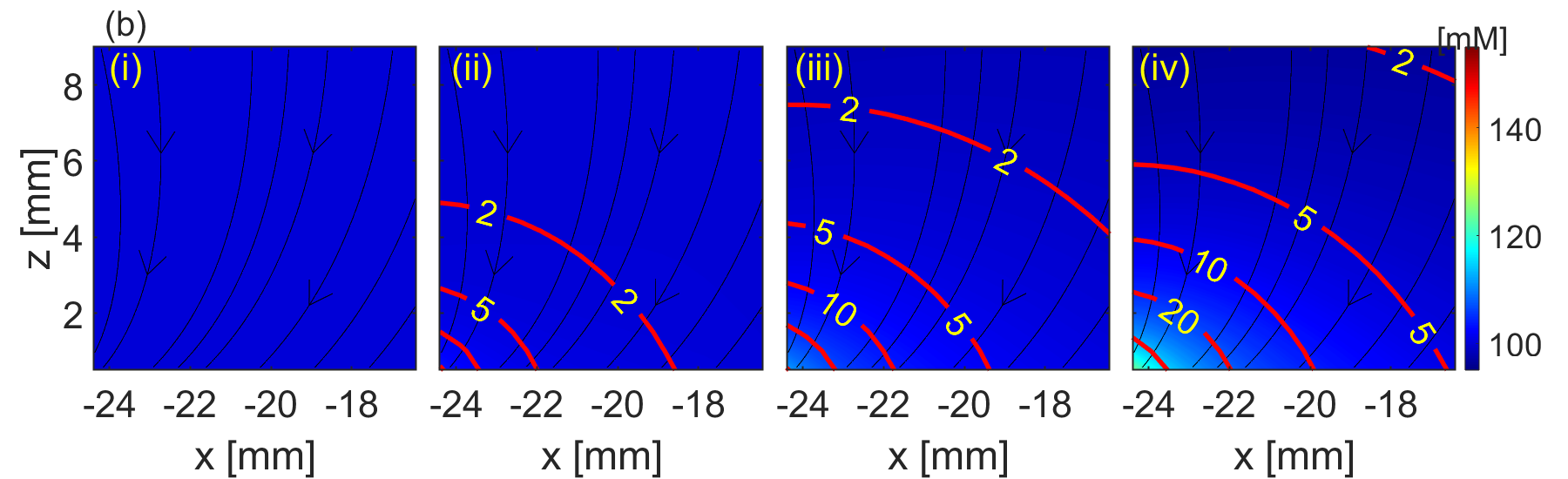}} \\
    \subfloat{\includegraphics[width=\linewidth]{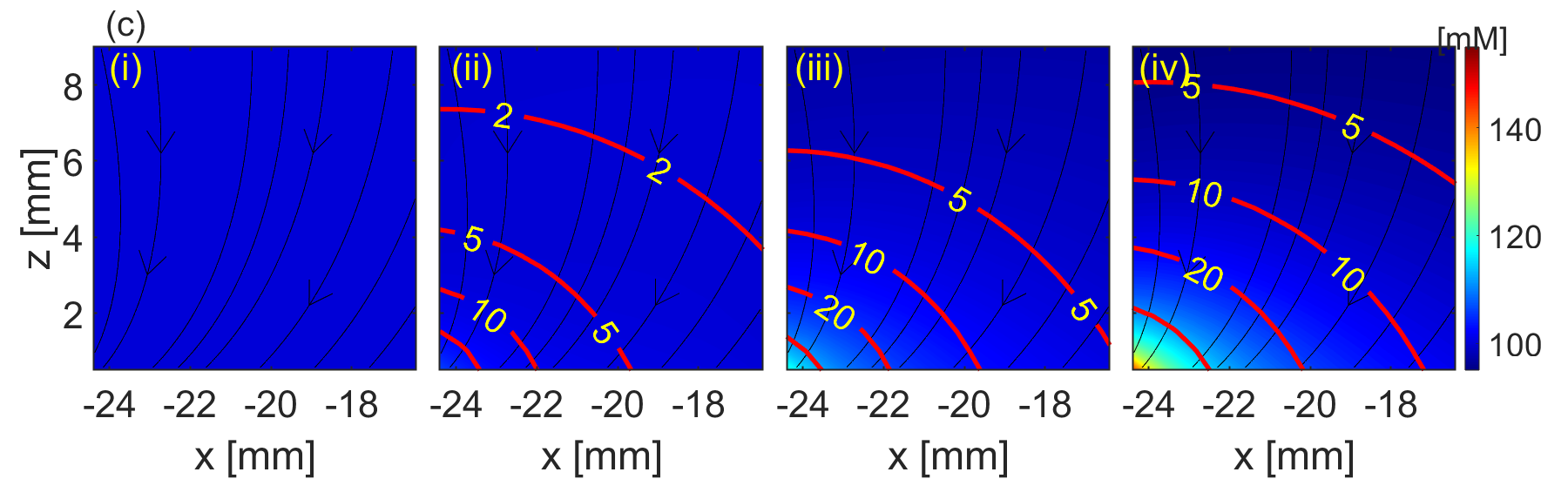}} \\
    \caption{Spatio-temporal evolution of MnCl$_2$ concentration in porous media with an initial concentration of 100~mM, simulated using Brinkman’s formulation. Panels correspond to particle sizes of (a) 63~$\mu$m, (b) 168~$\mu$m, and (c) 500~$\mu$m, respectively. Each subpanel presents concentration heatmaps overlaid with velocity vectors and contours at four time points: $t =$ 0~(i), 1~(ii), 12~(iii), and 72~(iv) hours. The color gradient represents ion enrichment toward the magnet, and velocities are expressed in nm/s.}
    \label{fig:MnCl2-diff-PorousMedia}
\end{figure}

\begin{figure}[H]
    \centering
    \subfloat{\includegraphics[width=\linewidth]{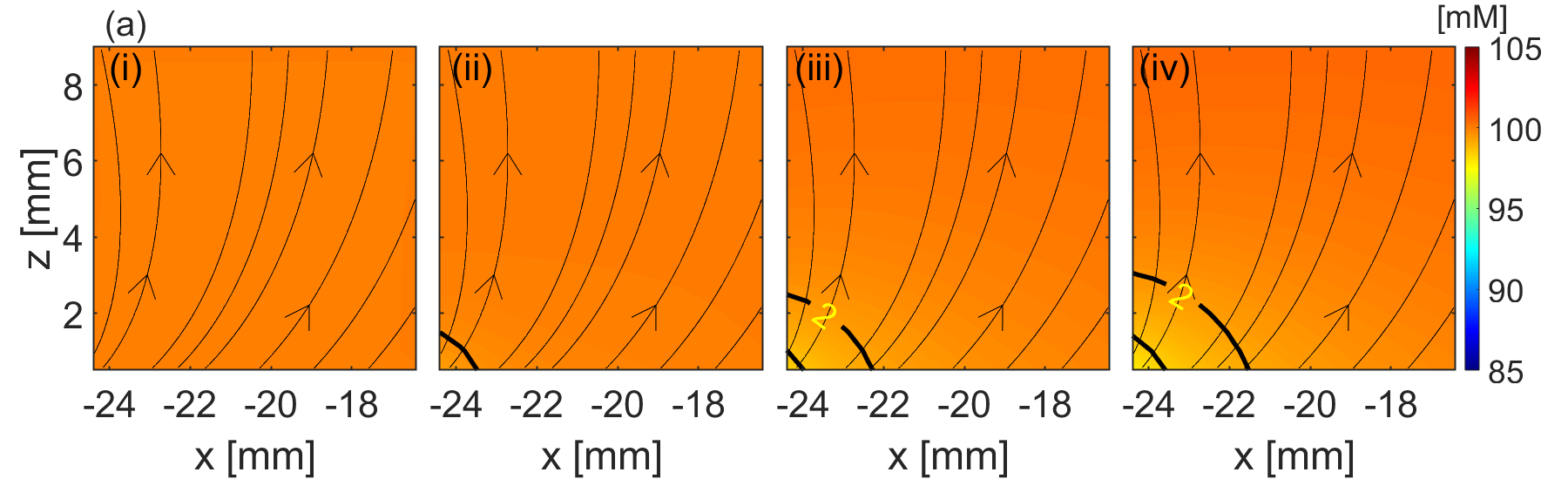}} \\
    \subfloat{\includegraphics[width=\linewidth]{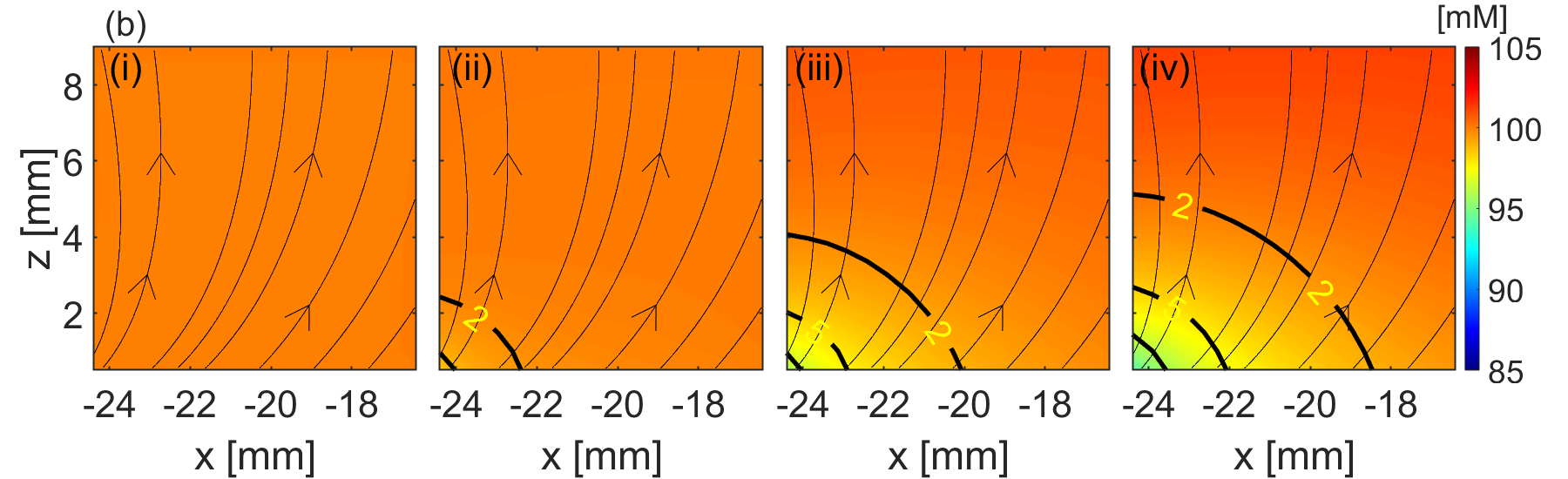}} \\
    \subfloat{\includegraphics[width=\linewidth]{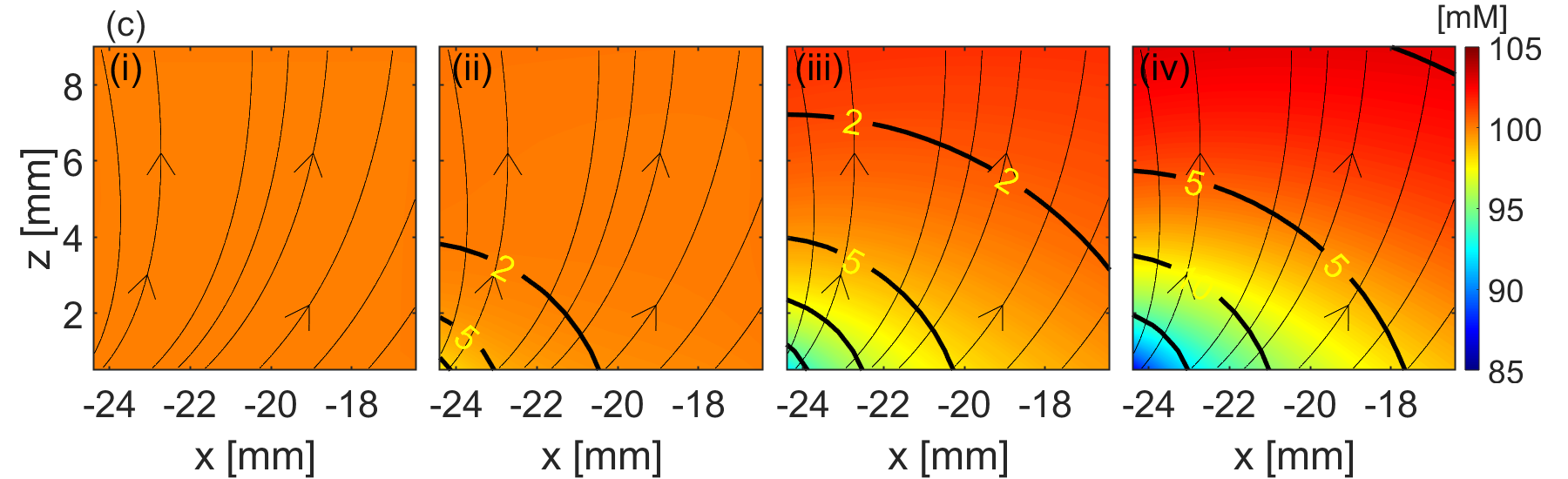}} \\
    \caption{Spatio-temporal evolution of ZnCl$_2$ concentration in porous media with an initial concentration of 100~mM, simulated using Brinkman’s formulation. Panels correspond to particle sizes of (a) 63~$\mu$m, (b) 168~$\mu$m, and (c) 500~$\mu$m, respectively. Each subpanel presents concentration heatmaps overlaid with velocity vectors and contours at four time points: $t =$ 0~(i), 1~(ii), 12~(iii), and 72~(iv) hours. The color gradient represents ion depletion toward the magnet, and velocities are expressed in nm/s.}
    \label{fig:ZnCl2-diff-PorousMedia}
\end{figure}


\begin{figure}[H]
    \centering
    \subfloat{{\includegraphics[width=0.6\linewidth]{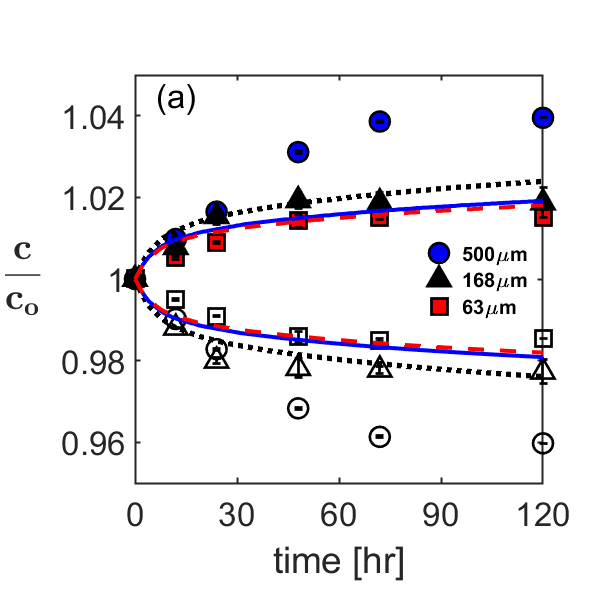}}}\\
  
    \caption{Temporal evolution of normalized MnCl$_2$ concentration in various porous media, along with predictions from the Brinkman formulation (solid curves) using permeability values estimated by the Carman-Kozeny model. In here the red, blue and black curves correspond to porous media with a 63 $\mu$m, 500 $\mu$m, and a polydisperse (denoted as 168 $\mu$m) particle size.}
    \label{Carman-Kozeny SI}
\end{figure}

\section*{Permeability Measurement Using Darcy’s Law}\label{SI Permeability section}

\begin{figure}[hthp]
    \centering
    \subfloat{{\includegraphics[width=0.7\linewidth]{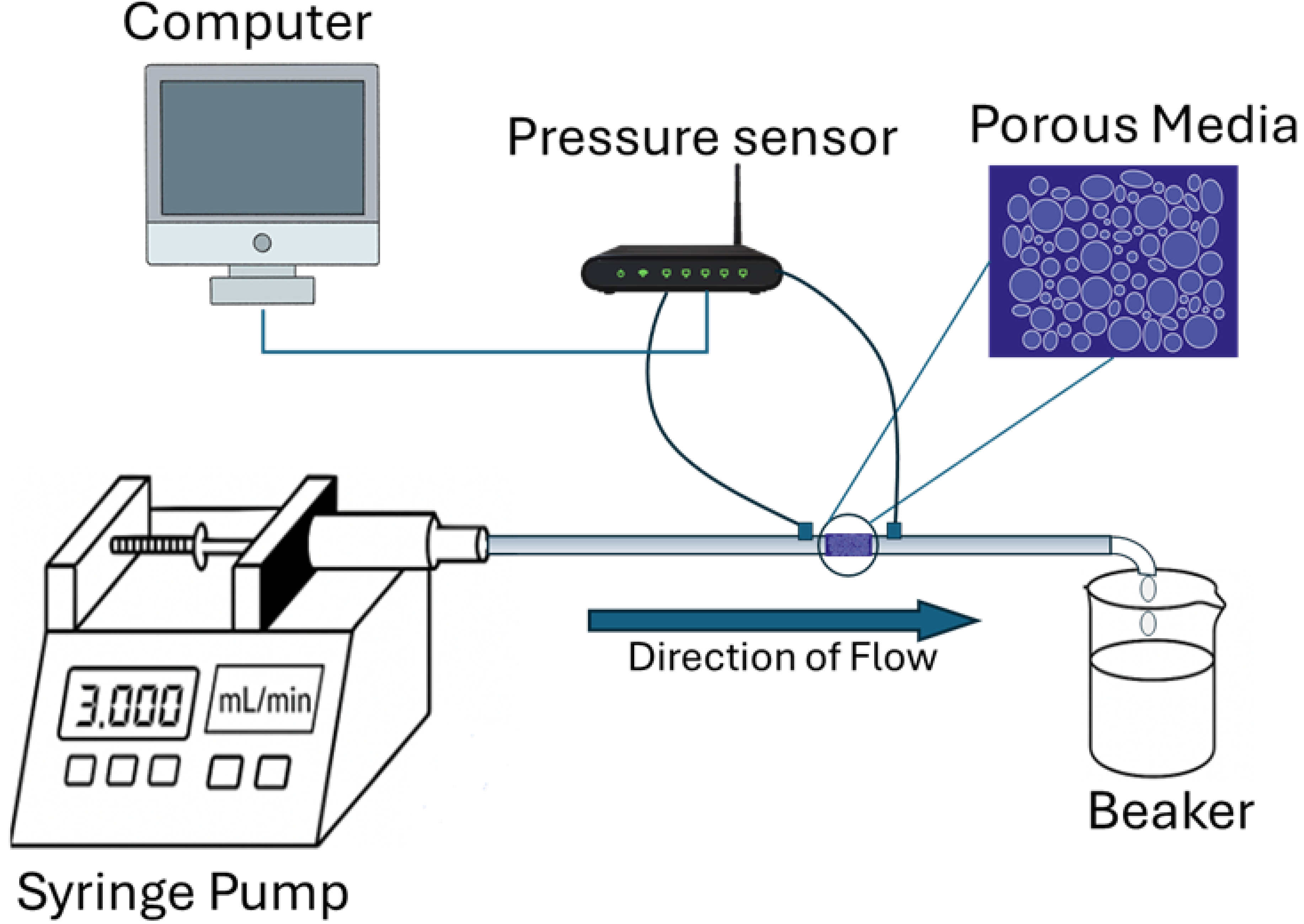}}}\\
    \caption{A schematic of the experimental setup used for permeability measurements of various porous media.}
    \label{experiment setup SI}
\end{figure}

The permeability of the porous media was determined using Darcy’s law, as shown in equation below using the experimental setup illustrated in Fig.~\ref{experiment setup SI}.

\begin{equation}\label{darcy}
    q = \frac{\kappa}{\mu L} \Delta P
\end{equation}

\noindent
where \( q \) is the volumetric flux [m$^3$/s], \( \kappa \) is the permeability of the porous medium [m²], \( \mu \) is the dynamic viscosity of the fluid [Pa·s], $L$ = 20 mm is the length of the porous section, and \( \Delta P \) is the pressure drop across the medium [Pa]. In this setup, a syringe pump delivers water to the inlet of a tube containing the porous medium, with the outlet discharging into a beaker. A differential pressure sensor is connected in parallel at both the inlet and the outlet of the porous section to record the pressure drop without disrupting the flow. To account for system resistance, an initial run was performed without the porous medium to quantify the pressure loss due to tubing and fittings. The value obtained was \(-188\) Pa, which was subtracted from all subsequent readings with the porous media in place.

\begin{figure}[hthp]
    \centering
    \subfloat{{\includegraphics[width=\linewidth]{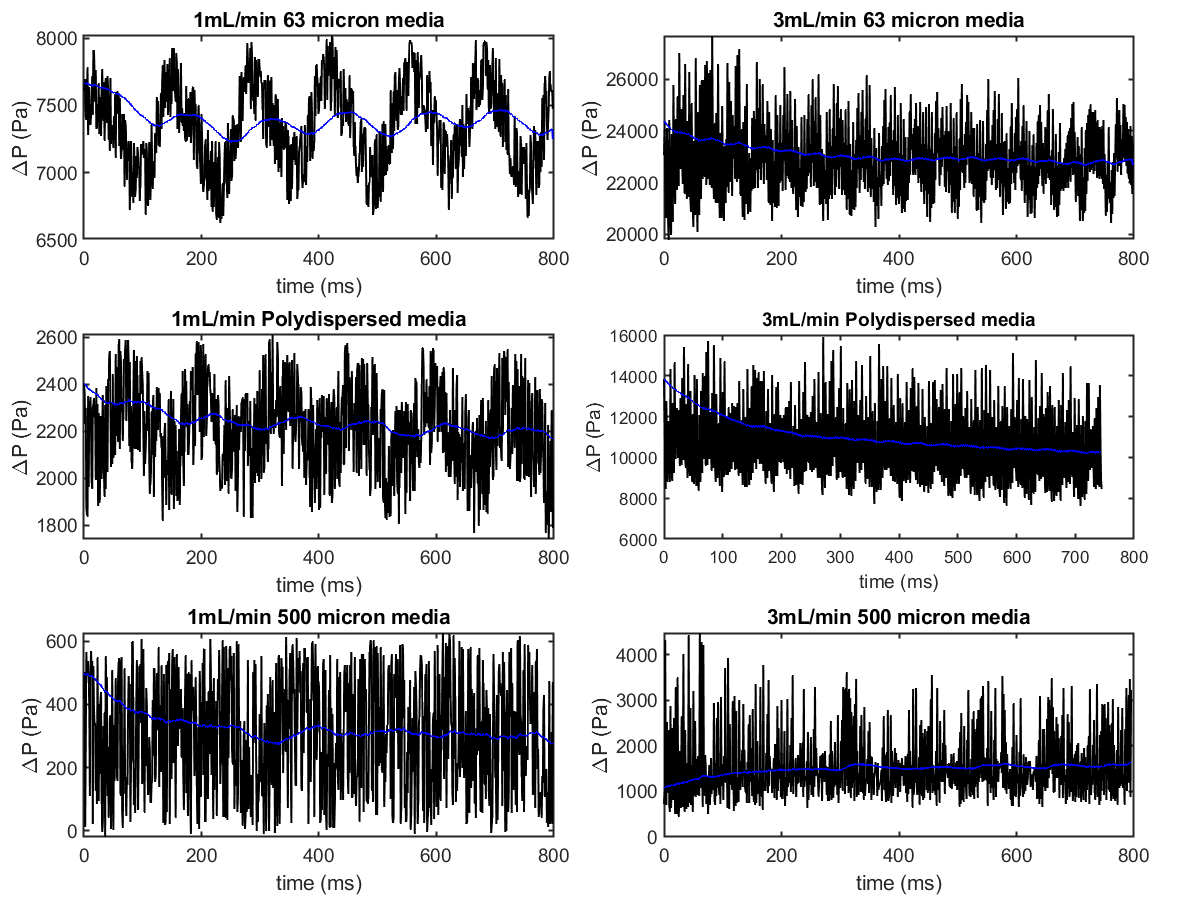}}}\\
    \caption{Pressure drop measurement for permeability in different porous media}
    \label{SI permeability experimental measurement}
\end{figure}

Measurements were then conducted for the three types of porous media described in the main text: (i) 63~$\mu$m monodispersed silica gel, (ii) polydispersed silica gel, and (iii) 500~$\mu$m monodispersed silica gel. For each case, pressure drops were recorded at two different volumetric flow rates: 1~mL/min and 3~mL/min. These flow rates enabled the evaluation of permeability under different flux conditions. The pressure drop values were obtained using computer software and are shown in black in Fig.~(\ref{SI permeability experimental measurement}). To determine the average pressure drop, the raw data were smoothed by averaging, resulting in a straight trend line displayed in blue. The corresponding permeability values, calculated from the averaged pressure drops, are presented in Table~\ref{experimental data SI}.

\begin{table}[hthp]
    \centering
    \caption{List of experimentally measured parameters for determining the permeability of the porous medium used in this study}
    \label{experimental data SI}
    \begin{tabular}{|c|c|c|c|}
    \hline
    \multirow{2}{*}{Porous media} & Flow rate  & Pressure drop  & Permeability   \\
     & [mL/min]  & $\Delta P$ [Pa] & $\kappa \times 10^{12}  [m^2] $ \\
    \hline
        \multirow{2}{*}{polydispersed}  & 1 & 2431 &  8.73 \\
        \cline{2-4}
        & 3 & 10742 &  5.93\\
        \hline
        \multirow{2}{*}{63 $\mu$m}  & 1 & 7581 &  2.80 \\
        \cline{2-4}
        & 3 & 24052 &  2.70\\
        \hline
        \multirow{2}{*}{500 $\mu$m}  & 1 & 517 &  41.08 \\
        \cline{2-4}
        & 3 & 1672 &  38.07\\
        \hline
    \end{tabular}
    
\end{table}

\end{document}